\documentclass[article,aps,pra,superscriptaddress,reprint,a4paper,longbibliography]{revtex4-2}

\usepackage{graphicx}
\usepackage{tikz}
\usepackage{amsmath,amsfonts,amssymb}
\usepackage{fullpage}
\usepackage[T1]{fontenc}
\usepackage{caption}
\usepackage{subcaption}
\usepackage{placeins}
\usepackage{xcolor}
\usepackage[colorinlistoftodos]{todonotes}
\usepackage{float}

\newcommand{\mbf}[1]{\mathbf{#1}}
\newcommand{\mrm}[1]{\mathrm{#1}}

\newcommand{\ka}{k}

\usepackage{dsfont}
\newcommand{\id}{\mathds{1}}
\usepackage{hyperref}

\usepackage{placeins}

\usepackage{bm}
\renewcommand{\vec}[1]{\bm{#1}}

\newcommand{\ex}{\vec{e}_x}
\newcommand{\ey}{\vec{e}_y}
\newcommand{\ez}{\vec{e}_z}

\usepackage{subcaption}

\definecolor{lime}{HTML}{A6CE39}
\DeclareRobustCommand{\orcidicon}{
	\begin{tikzpicture}
	\draw[lime, fill=lime] (0,0) 
	circle [radius=0.16] 
	node[white] {{\fontfamily{qag}\selectfont \tiny ID}};
	\draw[white, fill=white] (-0.0625,0.095) 
	circle [radius=0.007];
	\end{tikzpicture}
	\hspace{-2mm}
}
\foreach \x in {A, ..., Z}{\expandafter\xdef\csname orcid\x\endcsname{\noexpand\href{https://orcid.org/\csname orcidauthor\x\endcsname}
			{\noexpand\orcidicon}}
}

\begin{document}
\title{Zero-field optical magnetometer based on spin-alignment}

\author{A.~Meraki\orcidA{}}
\email{adil.meraki@nottingham.ac.uk}
\affiliation{School of Physics and Astronomy, University of Nottingham, University Park, Nottingham, NG7 2RD, UK}

\author{L.~Elson\orcidB{}}
\affiliation{School of Physics and Astronomy, University of Nottingham, University Park, Nottingham, NG7 2RD, UK}

\author{N.~Ho\orcidE{}}
\affiliation{School of Physics and Astronomy, University of Nottingham, University Park, Nottingham, NG7 2RD, UK}

\author{A.~Akbar\orcidC{}}
\affiliation{School of Physics and Astronomy, University of Nottingham, University Park, Nottingham, NG7 2RD, UK}

\author{M.~Ko\'{z}bia\l\orcidM{}}
\affiliation{Centre for Quantum Optical Technologies, Centre of New Technologies, University of Warsaw, Banacha 2c, 02-097 Warszawa, Poland}

\author{J.~Ko\l{}ody\'{n}ski\orcidJ{}}

\affiliation{Centre for Quantum Optical Technologies, Centre of New Technologies, University of Warsaw, Banacha 2c, 02-097 Warszawa, Poland}

\author{K.~Jensen\orcidD{}} 
\email{kasjensendk@gmail.com}
\affiliation{School of Physics and Astronomy, University of Nottingham, University Park, Nottingham, NG7 2RD, UK}

\begin{abstract}

Optically-pumped magnetometers are an important instrument for imaging biological magnetic signals without the need for cryogenic cooling. These magnetometers are presently available in the commercial market and utilize the principles of atomic alignment or orientation, enabling remarkable sensitivity and precision in the measurement of magnetic fields. This research focuses on utilizing a spin-aligned atomic ensemble for magnetometry at zero-field. A novel approach is introduced, which involves evaluating how the linear polarization of light rotates as it passes through the atomic vapor to null the magnetic field. Analytical expressions are derived for the resulting spin alignment and photodetection signals. Experimental results are provided, demonstrating good agreement with the theoretical predictions. The sensitivity and bandwidth of the magnetometer are characterized based on the detected polarization rotation signal. Lastly, the practical utility of the magnetometer for medical applications is demonstrated by successfully detecting a synthetic cardiac signal.
\end{abstract}

\maketitle

\section{Introduction} 

Over the past decade, significant progress has been made in enhancing the sensitivity of optically pumped magnetometers (OPMs) used in extremely low magnetic field environments \cite{budker2007optical,wilson2019ultrastable,kominis2003subfemtotesla}. These advancements have surpassed the sensitivity of superconducting quantum interference devices (SQUIDs) \cite{SQUIDReview} while eliminating the requirement for cryogenic cooling. Current commercially available OPMs \cite{Quspin,fieldline,twinleaf} are commonly operated in the spin-exchange relaxation-free (SERF) regime, enabling measurements of one or more components of the magnetic field near zero-field \cite{li2023parameter,le2022zero}. OPMs have demonstrated their exceptional ability to precisely measure small magnetic fields in a variety of applications, some examples include biomedical imaging \cite{jensen2018magnetocardiography, marmugi_renzoni_2016, brookes2022magnetoencephalography,xia2006magnetoencephalography,sander2012magnetoencephalography, wyllie2012optical}, metal detection \cite{rushton2022unshielded, bevington2019enhanced} and material characterizations \cite{romalis2011atomic}.  The miniaturization and commercialization of magnetometers is highly desired, especially for integrating them into densely packed arrays for medical imaging applications \cite{boto2018moving, borna2020non}. 

OPMs typically utilize optical pumping techniques with either circularly \cite{kominis2003subfemtotesla, rushton2022unshielded} or linearly polarized light \cite{weis2006theory, ledbetter2007detection,di2007sensitivity, rushton2023alignment}. Circularly polarized light is frequently used in most zero-field OPM configurations to orient the total spin of the atomic ensemble towards a specific direction \cite{savukov2005tunable}. 
Alternatively, one can use linearly polarized light, which enables the preparation of atomic alignment for spins larger than 1/2 \cite{zigdon2010nonlinear, ingleby2018vector}. Linearly polarized light in the alignment configuration can be used for pumping and probing atoms using a single laser beam. This can be utilized in a zero-field setting \cite{ Beato2018, Bertrand2021, wang2022all}. Zeroing the magnetic field in an OPM is crucial to minimize any unintended influences, such as ambient magnetic fields or equipment-induced disturbances. Effectively nulling the magnetic field allows researchers to set up a reliable reference point, enhancing the instrument's sensitivity and precision for a diverse range of applications.

In our work, we investigate how to use a spin-aligned atomic ensemble for zero-field magnetometry. We introduce a novel approach for nulling the magnetic field which involves evaluating how the  linear polarization of the light is rotated as it is transmitted through the atomic vapour. We derive analytical expressions for the generated spin-alignment and the photodetection signals. A room temperature paraffin coated cell containing caesium (Cs) is used to study how the polarization rotation signal changes with small magnetic fields near zero. By analyzing the polarization of light as it passes through the vapor cell under different magnetic field conditions, we determine the appropriate direction to adjust the compensation magnetic field and whether to increase or decrease its magnitude. 
Having presented the method for nulling the field, we then characterize the magnetometer's sensitivity and bandwidth from the detected polarization rotation signal. Our zero-field magnetometer is modulation-free, as opposed to previous work which typically employs oscillating magnetic fields for modulating the atomic response and where the absorption of light is detected \cite{Bertrand2021, Beato2018}. We finally demonstrate that our magnetometer is useful for medical applications by detecting a synthetic cardiac signal. 

\section{Theory}
\subsection{Equation of motion for alignment multipoles}
Our experiments are carried out using an ensemble of room-temperature caesium atoms in a vapour cell. Caesium has two ground states with hyperfine quantum numbers $F=3$ and $F=4$ and the first excited state has hyperfine quantum numbers $F'=3$ and $F'=4$ \cite{steck}. We use linearly polarized light on-resonant with the 
$6S_{1/2}$ $\rightarrow$ $6P_{1/2}$, $F=4 \rightarrow F'{=}3$ D1 transition at a wavelength of 895~nm 
to optically pump and probe the atoms. One can describe a single atom in the ground state with a particular hyperfine quantum number $F$ with a density matrix $\rho$ with size $\left(2F+1\right) \times \left(2F+1\right) $ using the basis $\left\{ m=-F, m=-F+1, ..., m=F\right\}$, where $m$ is the projection of the total angular momentum along the $\ez$ direction. Alternatively, one can express the quantum state as
\begin{equation}
\rho = \sum_{\ka=0}^{2F} \sum_{q=-\ka}^{\ka} m_{\ka,q} T_{q}^{k},
\end{equation}
where $m_{\ka,q}$ are multipoles with rank $\ka=0,1,...,2F$ and $q=-\ka,...,\ka$  and $T_q^{\ka}$ are irreducible spherical tensor operators \cite{weis2006theory, auzinsh_budker_rochester_2014, Beato2018}.
Note that in total there are $\sum_{\ka=0}^{2F} \left(2\ka+1 \right)=\left(2F+1\right)^2$ multipole components  which is equal to the number of density matrix elements. 
For a normalized state $m_{0,0}=1/\sqrt{2F+1}$.
The $\ka=1$ multipole describes how the spin is oriented, and they can be organised in a vector
\begin{equation}
\mbf{m}_1 = \left(m_{1,-1}, m_{1,0}, m_{1,1}\right)^T.
\end{equation}
Here ``T'' denotes the transpose such that $\mbf{m}_1$ is a 3-dimensional column vector.
The $\ka=2$  multipoles describe how the spin is aligned along some axes, and they can be organised in the 5-dimensional column vector
\begin{equation}
\mbf{m}_2 = \left(m_{2,-2}, m_{2,-1}, m_{2,0}, m_{2,1}, m_{2,2} \right)^T.
\end{equation}
It can be convenient to describe the quantum state using the multipoles instead of the density matrix, due to the properties of the multipoles under rotations. For example, atoms optically pumped with circular polarized light can be well described by their orientation $\mbf{m}_1$, while atoms optically pumped with linear polarized light can be well described by their alignment $\mbf{m}_2$.

\begin{figure}
    \centering    
    \begin{subfigure}[b]{0.44\linewidth}
        \centering
        \includegraphics[width=\linewidth]{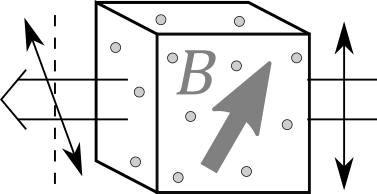}
        \caption{}
        \label{fig:idea1}
    \end{subfigure}
    \hfill
    \begin{subfigure}[b]{0.54\linewidth}
        \centering
        \includegraphics[width=\linewidth]{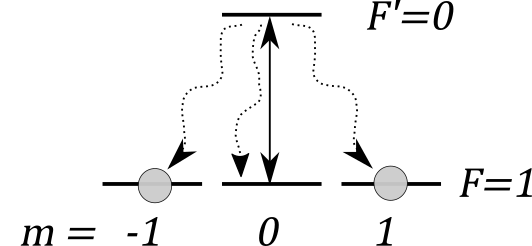}
        \caption{}
        \label{fig:idea2}
    \end{subfigure}    
     \hfill
    \begin{subfigure}[b]{0.99\linewidth}
        \centering
        \includegraphics[width=\linewidth]{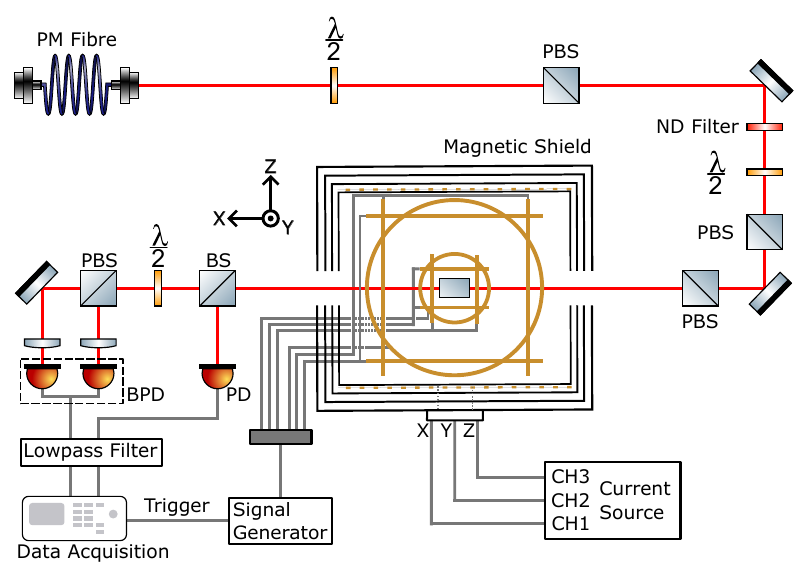}
        \caption{}
        \label{fig:setup} 
    \end{subfigure} 
    \caption{(a) Sketch of the polarization rotation of light transmitted through an atomic ensemble. (b) Toy model of an atom optically pumped with linearly polarized light. (c) Experimental setup. Red paths indicate optical pathways and grey paths represent the routes for the electrical signals. PM Fiber, polarization maintaining fiber; PBS, polarizing beam splitter; $\lambda /2$, half-wave plates; BPD, balanced photodetector.}
\end{figure}

For simplicity, we will use a simpler toy-model to describe the evolution of the single atom quantum state. Consider an atom with a $F=1$ ground state that has three sub-levels $\left\{ m=-1, m=0, m=1\right\}$ and a single excited state with $F'=0,m=0$, where the quantization axis is chosen along the $\ez$ direction.  The $F=1$ ground state can be described either by a $3 \times 3$ density matrix $\rho$ or by the multipoles $m_{\ka,q}$ with $\ka=0,1,2$ and $q=-\ka,...,\ka$. 

Let us assume that the beam of light, linearly polarized along the $\ez$ direction, is propagating through an ensemble of atoms, as illustrated in Fig.~\ref{fig:idea1}. In the absence of any magnetic field, the light will optically pump and polarize each atom. With perfect optical pumping, where the populations in the $m=1$ and $m=-1$ sub-levels are both equal to $1/2$, as illustrated in Fig.~\ref{fig:idea2}, we have $m_{0,0}=1/\sqrt{3}$, 
$\mbf{m}_1 = \left(0,0,0 \right)^T$
and
$\mbf{m}_2 = \left( 0,0,1/\sqrt{6},0,0 \right)^T$.
We see that the orientation multipole is zero and that the quantum state is described with just the alignment multipole.
If the optical pumping is not perfect, then the generated alignment will be reduced. In any case, the generated alignment can be described by the expression
\begin{equation}
\mathbf{m}_2^{\mrm{eq}} = \left(0,0, m_{2,0}^{\mrm{eq}}, 0, 0 \right)^T.
\end{equation}

We now assume the atom is placed in a magnetic field
\begin{equation}
\vec{B} = B_x \ex + B_y \ey + B_z \ez .
\end{equation}
Due to the properties of spherical tensors under rotations, it is possible to rewrite the Liouville equation for the density matrix as a matrix equation for the multipoles. Including decay and repopulation terms, we find a matrix equation written compactly as
\begin{align} \label{eq:m2Bxyz}
\dot{\mbf{m}}_2 = &
\left(
-i \omega_x J_x^{(2)}
-i \omega_y J_y^{(2)}
-i \omega_z J_z^{(2)} 
\right) \mbf{m}_2 \nonumber \\
& -\Gamma \id_5 + \Gamma  \mbf{m}_2^{\mrm{eq}}.
\end{align}
Here $\omega_x=\gamma B_x$, $\omega_y=\gamma B_y$, and $\omega_z=\gamma B_z$, where $\gamma$ is the gyromagnetic ratio of the atom.
It is possible to extend our model to include multiple relaxation rates \cite{weis2006theory, di2007sensitivity}. However, for simplicity, we assume just a single ground-state relaxation rate $\Gamma$.
The matrices $J_x^{(2)}$, $J_y^{(2)}$, and $J_z^{(2)}$ are the $\ka=2$ representation of the angular momentum operators and $\id_5$ is the $5 \times 5$ identity matrix. 
Eq.~(\ref{eq:m2Bxyz}) has been explicitly written out in Appendix~A in the Supplemental Material. 
We can find the steady-state solution to Eq~(\ref{eq:m2Bxyz}) by setting the time-derivative to zero $\dot{\mbf{m}}_2^{\mrm{ss}} = 0$ and solve the system of linear equations. 
We find
\begin{widetext}
\begin{align} \label{eq:zerofieldss}
\begin{pmatrix} m_{2,-2}^{\mrm{ss}}\\m_{2,-1}^{\mrm{ss}}\\m_{2,0}^{\mrm{ss}}\\m_{2,1}^{\mrm{ss}}\\m_{2,2}^{\mrm{ss}}\end{pmatrix}
= &
\frac{m_{2,0}^{\mrm{eq}}}{\left( \Gamma^2 + \omega_x^2 + \omega_y^2+ \omega_z^2 \right) 
\left[ \Gamma^2 + 4\left( \omega_x^2 + \omega_y^2+ \omega_z^2 \right) \right]}
\nonumber \\
 \times &
\begin{pmatrix} 
-\sqrt{\frac{3}{2}} \left( \omega_x + i \omega_y \right)^2
\left( \Gamma^2 + \omega_x^2 + \omega_y^2 +3 i \Gamma \omega_z -2 \omega_z^2 \right) 
\\
-\sqrt{\frac{3}{2}} \left( \omega_x + i \omega_y \right) 
\left(i \Gamma + 2 \omega_z \right)
\left( \Gamma^2 + \omega_x^2 + \omega_y^2 +3 i \Gamma \omega_z -2 \omega_z^2 \right) 
\\
\Gamma^4 + \left( \omega_x^2 + \omega_y^2 - 2 \omega_z^2 \right)^2
+ \Gamma^2 \left[ 2 \left( \omega_x^2 + \omega_y^2 \right) + 5 \omega_z^2 \right]
\\
\sqrt{\frac{3}{2}} \left( \omega_x - i \omega_y \right) 
\left(-i \Gamma + 2 \omega_z \right)
\left( \Gamma^2 + \omega_x^2 + \omega_y^2 - 3 i \Gamma \omega_z -2 \omega_z^2 \right) 
\\
-\sqrt{\frac{3}{2}} \left( \omega_x - i \omega_y \right)^2
\left( \Gamma^2 + \omega_x^2 + \omega_y^2 - 3 i \Gamma \omega_z -2 \omega_z^2 \right) 
\end{pmatrix}.
\end{align}
\end{widetext}
Note that $\left( m_{2,-2}^{\mrm{ss}} \right)^* = m_{2,2}^{\mrm{ss}}$ and
$\left( m_{2,-1}^{\mrm{ss}} \right)^*  = -m_{2,1}^{\mrm{ss}}$, where 
$^*$ denotes the complex conjugate. We also see that $m_{2,0}^{\mrm{ss}}=m_{2,0}^{\mrm{eq}}$ when the field is zero ($\omega_x = \omega_y = \omega_z$).

\subsection{Magnetometry signals}
We can calculate the observable signals from the steady-state multipoles in Eq.~(\ref{eq:zerofieldss}). 
We first consider a measurement of how much light is transmitted through a vapour cell with length $l$. We have $I(l)=I(0) e^{-\kappa l}$, where $I(0)$ is the light intensity before the vapour cell, $I(l)$ is the light intensity after the vapour cell, and $\kappa$ is the absorption coefficient. The transmitted light can be measured with a single photodetector. Alternatively, we can measure how the linear polarization of the light is rotated as it traverses the vapour cell. This polarization rotation signal $S$ can be measured using a balanced photodetector.
One can calculate formulas for the absorption $\kappa$ and the polarization rotation signal $S$. Such formulas depend on how the coordinate system is defined. 
Above we assumed that the light is polarized in the $\ez$ direction and we now furthermore assume that the light is propagating in the $\ex$ direction. In this case, one can show (see Appendix~B in the Supplemental Material) that the absorption and polarization-rotation signals generally obey
\begin{equation} \label{eq:kappa}
\kappa \propto \left(m_{0,0}-\chi\cdot m_{2,0}\right),
\end{equation}
and
\begin{equation} \label{eq:S}
S \propto  i\left( m_{2,1}+m_{2,-1} \right),
\end{equation}
respectively, where only the parameter $\chi$ depends on the hyperfine quantum number $F$ considered.

For instance, in the case of $F=1$ we have that $\chi=\sqrt{2}$ and  $m_{0,0}=1/\sqrt{3}$ (for a normalized state).
For an un-polarized state, where each $m$-sublevel is occupied with $1/3$ probability, all multipoles are zero except $m_{0,0}$.
Therefore, the absorption for an un-polarized state will be 
$\kappa \propto m_{0,0}=1/\sqrt{3}$.
On the other hand, for a fully polarized state, the absorption will be $\kappa \propto \left(m_{0,0}-\sqrt{2} m_{2,0}\right) 
= \frac{1}{\sqrt{3}} - \sqrt{2} \cdot \frac{1}{\sqrt{6}} =  0$ at zero-field. I.e., the absorption is zero, and all light is transmitted. This makes sense as the atoms are optically pumped into a dark state.
For a non-zero magnetic field, we insert the steady state solution given by Eq.~(\ref{eq:zerofieldss}) into Eqs.~(\ref{eq:kappa}) and (\ref{eq:S}) and obtain
\begin{widetext}
\begin{equation} \label{eq:kappa2}
\kappa \propto m_{0,0} - \chi \cdot \frac{ m_{2,0}^{\mrm{eq}} 
\left[ 
\Gamma^4 +\left( \omega_x^2 + \omega_y^2 -2 \omega_z^2\right)^2 + \Gamma^2 \left\{ 2 \left( \omega_x^2 + \omega_y^2 \right)  + 5 \omega_z^2 \right\} 
\right]
}{
\left( \Gamma^2 +  \omega_x^2 + \omega_y^2 + \omega_z^2  \right)  \left[ \Gamma^2 + 4 \left( \omega_x^2 + \omega_y^2 + \omega_z^2  \right) \right] 
},
\end{equation}
and
\begin{equation} \label{eq:Sy_ex}
S \propto \frac{\sqrt{6} m_{2,0}^{\mrm{eq}} \left[ \Gamma^3 \omega_x - \Gamma^2 \omega_y \omega_z + 2 \omega_y \omega_z \left( \omega_x^2 + \omega_y^2 -2 \omega_z^2\right) + \Gamma \omega_x \left( \omega_x^2 + \omega_y^2 + 4 \omega_z^2 \right) \right]}{\left( \Gamma^2 +  \omega_x^2 + \omega_y^2 + \omega_z^2  \right)   \left[ \Gamma^2 + 4 \left( \omega_x^2 + \omega_y^2 + \omega_z^2  \right) \right] }.
\end{equation}
\end{widetext}

If we assume two components of the magnetic field are zero (e.g. $\omega_y$ and $\omega_z$) and that the third component is large (e.g. $\omega_x \gg \Gamma$), then Eq.~(\ref{eq:kappa2}) simplifies to 
\begin{equation}
\kappa \propto m_{0,0} - \chi \cdot 
m_{2,0}^{\mrm{eq}} \cdot \frac{1}{4}.
\end{equation}
This means that the absorption in this case is $1/4$ compared to that of an un-polarized state.

We finally emphasize that in the theoretical description of our magnetometer we assumed a toy model atom with a $F=1$ ground state, which obviously is an approximation for a real cesium atom which have hyperfine structure and two ground states with $F=3$ and $F=4$. We note that all numbered equations are still valid for a cesium atom, if we assume that the multipoles correspond to the density matrix in the $F=4$ sub-space.  Although a state with $F=4$ in general can have multipoles with $\ka=0,1,...,8$, only the $\ka=2$ multipoles are relevant for our work. This is because the light only probes the $\ka \leq 2$ multipoles (when assuming the light power is low), and because multipoles with different $\ka$'s do not couple to each other under rotations (e.g. during the evolution in a magnetic field). For $F=4$ we have $\chi=\tfrac{1}{2}\sqrt{\tfrac{11}{7}}$ which can be calculated from Eq.~(\ref{eq:kappa}) by noting that the state
$\rho = \left(1/2 \right) |F=4,m=4 \rangle \langle F=4,m=4 | 
+ \left(1/2 \right) |F=4,m=-4 \rangle \langle F=4,m=-4 | 
$
is a dark state for light linearly polarized along $\ez$.

\subsection{Magnetometry near zero-field}
It is instructive to Taylor expand the full formulas for the absorption and polarization rotation given by  Eqs.~(\ref{eq:kappa2}) and (\ref{eq:Sy_ex}) to obtain simpler expressions for the signal. To keep notation short, we define $x=\omega_x/\Gamma$, $y=\omega_y/\Gamma$, and $z=\omega_z/\Gamma$ which are proportional to the $B_x$, $B_y$, and $B_z$ components of the magnetic field, respectively.

We first consider the absorption measurement.
For example, if sweeping the $B_x$ magnetic field, and assuming the $B_y$ and $B_z$ are close to zero 
(i.e. $y,z \ll 1$), then we find 
\begin{equation} \label{eq:kappa_sweepBx}
\kappa \propto m_{0,0} - \chi \cdot  m_{2,0}^{\mrm{eq}} 
\cdot
\left[
\frac{1}{1+x^2}
\right].
\end{equation}
I.e., the absorption signal has a Lorentzian lineshape as a function of $x$, and the absorption does not depend on $y$ and $z$ to first order.
Similarly, if sweeping the $B_y$ magnetic field, and assuming the $B_x$ and $B_z$ are close to zero 
(i.e. $x,z \ll 1$), then we find 
\begin{equation}  \label{eq:kappa_sweepBy}
\kappa \propto m_{0,0} - \chi \cdot m_{2,0}^{\mrm{eq}} 
\cdot
\left[
\frac{1}{1+y^2}
\right].
\end{equation}
Finally, if sweeping the $B_z$ magnetic field, and assuming the $B_x$ and $B_y$ are close to zero 
(i.e. $x,y \ll 1$), then we find 
\begin{equation}  \label{eq:kappa_sweepBz}
\kappa \propto m_{0,0} - \chi \cdot m_{2,0}^{\mrm{eq}} 
\cdot 1.
\end{equation}
In this case the absorption is constant and independent of $x$, $y$ and $z$ to first order.

We now consider the polarization rotation signal.
If sweeping the $B_x$ magnetic field, and assuming the $B_y$ and $B_z$ are close to zero 
(i.e. $y,z \ll 1$), then we find to first order in $y$ and $z$ that
\begin{equation} \label{eq:S_sweepBx}
S \propto  \sqrt{6} m_{2,0}^{\mrm{eq}}
\cdot
\left[
\frac{x}{1+4x^2}
\right].
\end{equation}
This implies that signal does not depend on $y$ and $z$ to the first order and it has a dispersive lineshape as a function of $x$.
We can also calculate the slope (i.e. the derivative with respect to $x$) of Eq.~(\ref{eq:Sy_ex}) to second order in $y$ and $z$. We find
\begin{equation} \label{eq:S_sweepBx_slope}
\frac{\partial S}{\partial x}_{|x=0}
\propto  
\sqrt{6} m_{2,0}^{\mrm{eq}} \cdot
\left[
\frac{1+y^2+4z^2 }{ \left( 1+y^2+z^2 \right)  \left( 1+4y^2+4z^2 \right) }
\right],
\end{equation}
which means that the slope is maximal when $y=z=0$.

On the other hand, if sweeping the $B_y$ magnetic field, and assuming the $B_x$ and $B_z$ are close to zero 
(i.e. $x,z \ll 1$), then we find to first order in $x$ and $z$ that
\begin{equation} \label{eq:S_sweepBy}
S \propto  \sqrt{6} m_{2,0}^{\mrm{eq}} \cdot
\left[
x \cdot \frac{1}{1+4y^2} + 
z \cdot \frac{y \left(-1+2 y^2 \right) }{\left(1+y^2 \right) \left(1+4y^2 \right)}
\right].
\end{equation}
Finally, if sweeping the $B_z$ magnetic field, and assuming the $B_x$ and $B_y$ are close to zero 
(i.e. $x,y \ll 1$), then we find to first order in $x$ and $y$ that
\begin{equation} \label{eq:S_sweepBz}
S \propto  \sqrt{6} m_{2,0}^{\mrm{eq}} \cdot
\left[
x \cdot \frac{1}{1+z^2} -
y \cdot \frac{z}{1+z^2}
\right].
\end{equation}
From Eqs.~(\ref{eq:S_sweepBx}), (\ref{eq:S_sweepBx_slope}) (\ref{eq:S_sweepBy}), and (\ref{eq:S_sweepBz}) we see that the polarization rotation signal to first order is a measure of the $B_x$ component of the field, i.e. the component along the light propagation direction.

We also see that it is possible to use the polarization rotation signal given by Eqs.~(\ref{eq:S_sweepBy}) and (\ref{eq:S_sweepBz}) as a sensitive method for nulling the magnetic field. Besides the measured polarization rotation signal, one needs a coil system for applying fields in the $\ex$, $\ey$ and $\ez$ directions. For example, as seen from Eq.~(\ref{eq:S_sweepBy}), if sweeping the $B_y$ magnetic field, one can adjust the $B_x$ and $B_z$ magnetic field components using the coil system until the signal is zero. When that is the case, those field components will be nulled $B_x=B_z=0$. 
Similarly, as seen from Eq.~(\ref{eq:S_sweepBz}), if sweeping the $B_z$ magnetic field, one can adjust the $B_x$ and $B_y$ magnetic field components until the signal is zero, and when that is the case, $B_x=B_y=0$.

\section{Experimental Setup and Procedure}

We implemented the zero-field magnetometry based on spin-alignment to a table-top, laboratory setup, see schematics in Fig.~\ref{fig:setup}.  Light with wavelength 895~nm resonant with the $F=4\rightarrow F'=3$ Cs D1 transition was emitted from a fiber pigtailed diode laser (Thorlabs DBR895PN). To adjust the intensity of light, a half-wave plate was employed in combination with a polarizer that aligned the polarization along the $\ez$-axis. 
The light power (measured before the vapour cell) was 3~$\mu$W for the data presented in Figs.~\ref{fig:TE_sweep_z},  \ref{fig:TE_sweep_y}, \ref{fig:Hanle}, and \ref{fig:TE_sweep_x}, 
and 10~$\mu$W for the data presented in Figs.~
\ref{fig:bandwidth} and
\ref{fig:Individual Cardiac Pulse}.
The linearly $z$-polarized light entered a  $\left(5~\mathrm{mm}\right)^3$ cubic vapor cell 
which was located in a four-layer mu-metal magnetic shield (Twinleaf MS-1L). The vapor cell is hand-blown, contains caesium atomic vapour, and is coated with paraffin. This leads to a long spin-coherence time.
Following that, polarimetry is performed by employing a half-wave plate, a polarizing beam splitter and a balanced photodetector (Thorlabs PDB210A/M). The output from the balanced photodetector is subsequently directed through a 2~kHz low-pass filter and recorded with a data acquisition card within the computer for further analysis.
For the absorption measurements (presented in Sec.~\ref{sec:exp-absorption}), the transmitted light intensity was measured directly with a single photodetector (Thorlabs PDA36A2) placed immediately after the vapour cell outside the magnetic shield.
The caesium cell is positioned within three sets of coils, where each coil set can be used to apply fields in three ($\ex$, $\ey$, $\ez$) directions.
With no current applied to the coils, there is a small residual magnetic field 
$\vec{B}^{\mathrm{residual}}$ inside the shield which is on the order of 100~nT.
The outer coil set (which comes as standard with the Twinleaf MS-1L magnetic shield) is used to apply a static field $\vec{B}^{\mathrm{DC}}$  to compensate the residual magnetic field in order for us to completely null the total magnetic field.
The middle coil set is a three-axis cubic Helmholtz coil set which can be used to generate a linearly sweeping magnetic field $ \vec{B}^{\mathrm{sweep}}$ which is used in the procedure for nulling the static field. 
The inner coil set, which is also a three-axis cubic Helmholtz coil, is used for generating different signals $\vec{B}^{\mathrm{signal}}$, e.g.\ a sinusoidal magnetic field for characterizing the magnetometer response and sensitivity or a synthetic cardiac signal.
The total magnetic field can be written as
\begin{equation}
\vec{B} =  
\vec{B}^{\mathrm{residual}} +
\vec{B}^{\mathrm{DC}} +
\vec{B}^{\mathrm{sweep}} +
\vec{B}^{\mathrm{signal}} .
\end{equation}

\section{Results and Discussions}

\subsection{Method for nulling the static field}

\begin{figure*}
    \centering
    \begin{subfigure}[b]{0.49\linewidth}
        \centering
        \includegraphics[width=\linewidth]{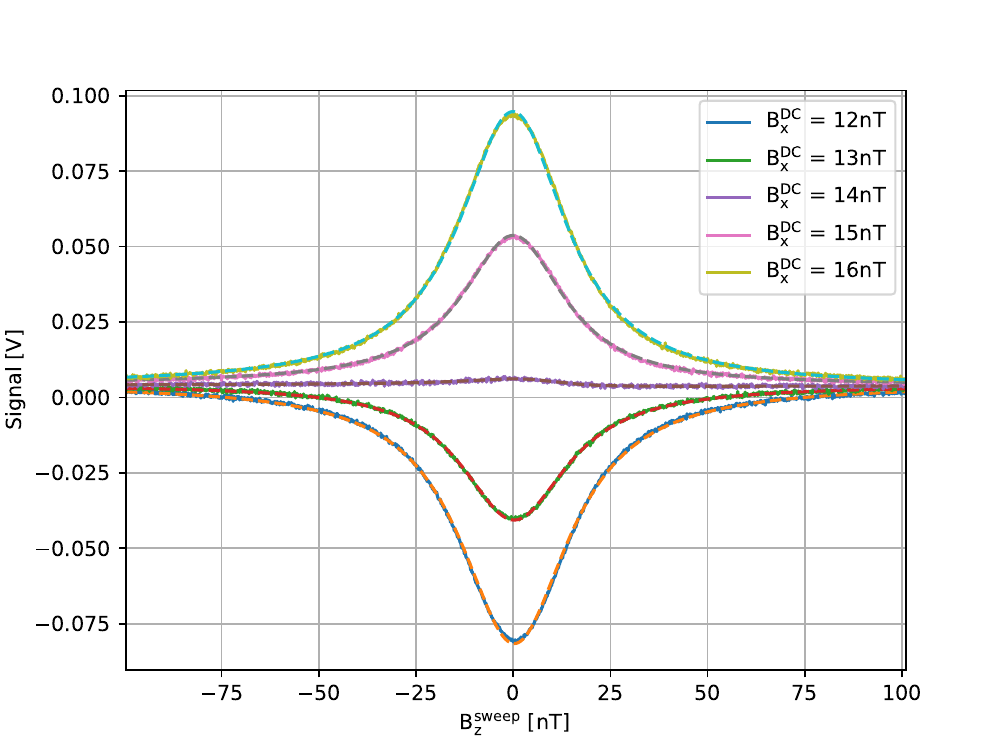}
        \caption{}
        \label{fig:sweep_z_vary_x}
    \end{subfigure}
    \hfill
    \begin{subfigure}[b]{0.49\linewidth}
        \centering
        \includegraphics[width=\linewidth]{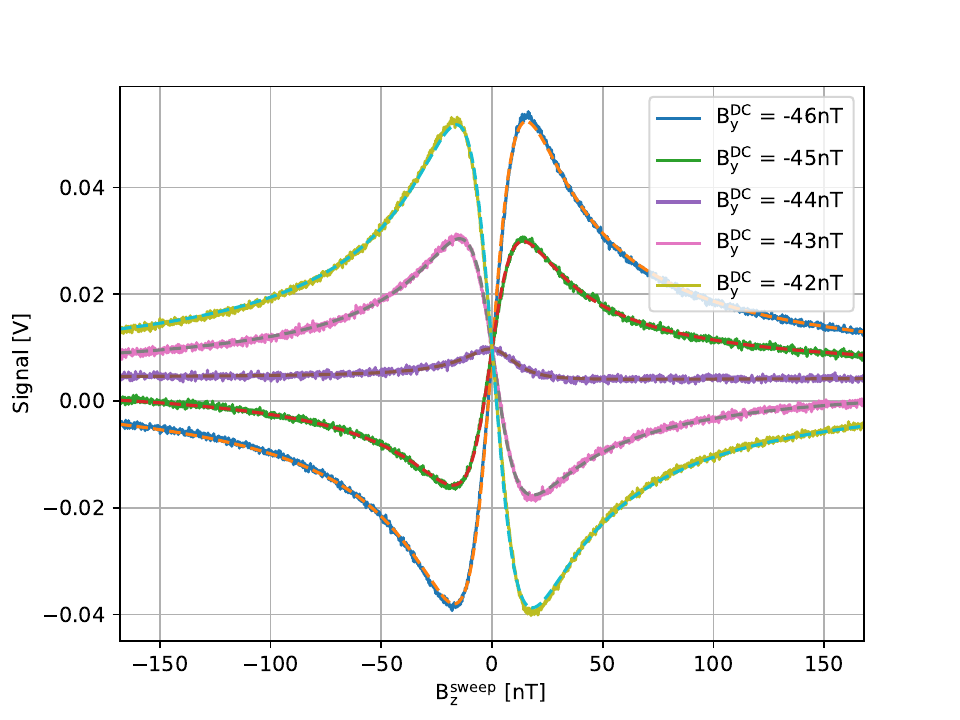}
        \caption{}
        \label{fig:sweep_z_vary_y}
    \end{subfigure}
    
    \vspace{0.3em}

    \begin{subfigure}[b]{0.49\linewidth}
        \centering
        \includegraphics[width=\linewidth]{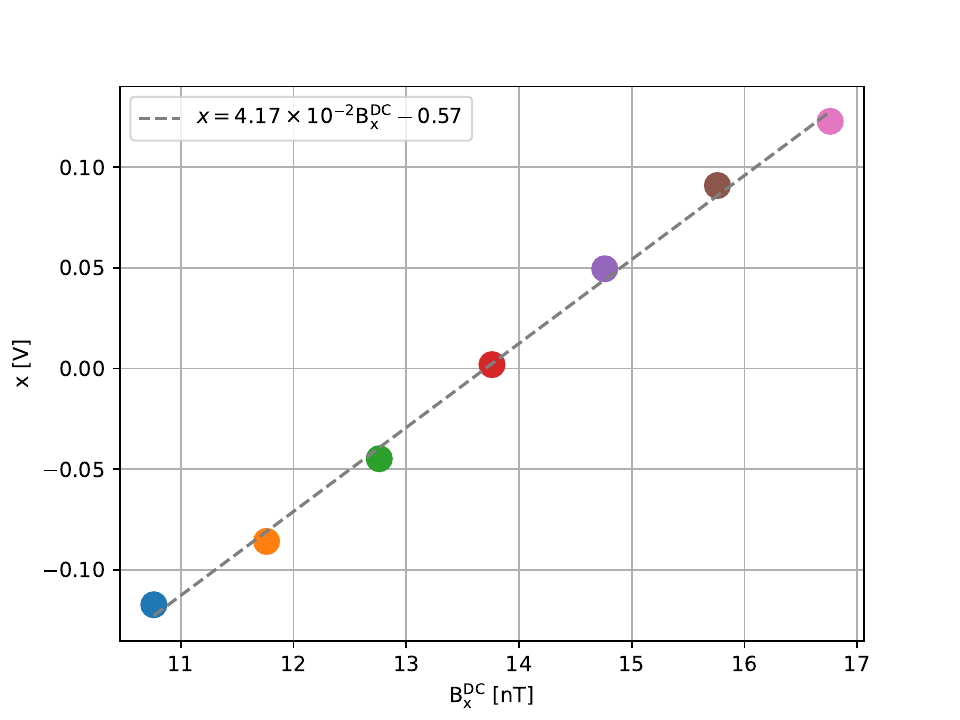}
        \caption{}
        \label{fig:sweep_z_vary_x_fit}
    \end{subfigure}
    \hfill
    \begin{subfigure}[b]{0.49\linewidth}
        \centering
        \includegraphics[width=\linewidth]{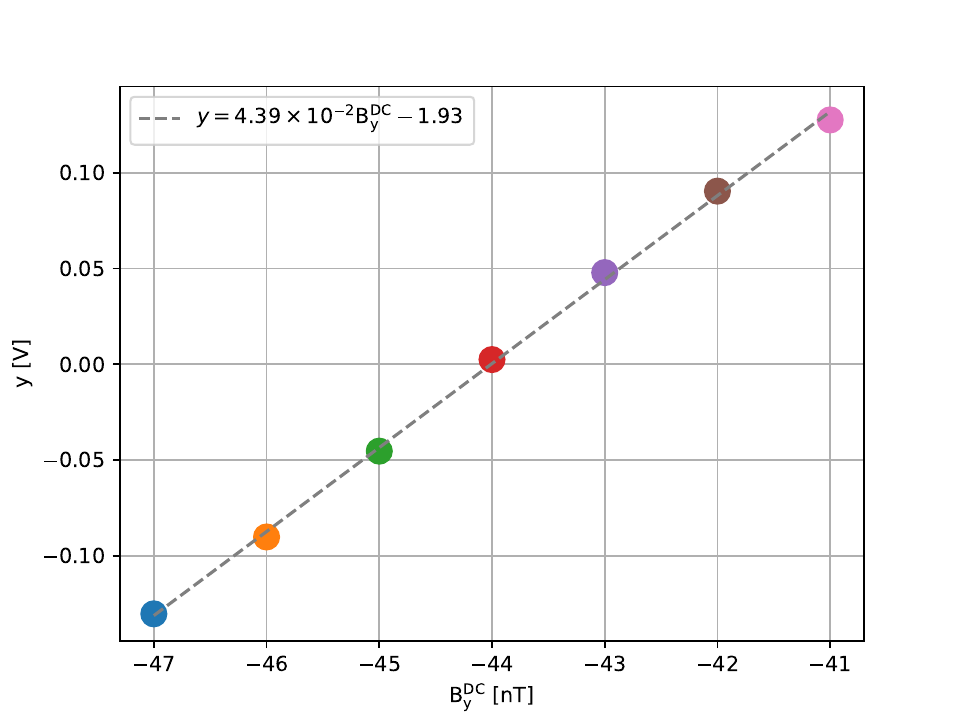}
        \caption{}
        \label{fig:sweep_z_vary_y_fit}
    \end{subfigure}
    
     \caption{Sweeping a magnetic field $B_z^{\mathrm{sweep}}$ applied along the $\ez$ direction near zero-field condition. Solid lines are experimental data, and dashed lines are fits to Eq.~(\ref{eq:S_sweepBz}). (a) and (b) show signals for various settings of magnetic field applied along the $\ex$ and $\ey$ directions, respectively. 
    (c) and (d) show the corresponding fit parameters  
     $x\propto \left( B_x^{\mathrm{residual}} + B_x^{\mathrm{DC}} \right)$
    and 
     $y\propto \left( B_y^{\mathrm{residual}} + B_y^{\mathrm{DC}} \right)$. The DC fields are  $B_z^{\mathrm{DC}} = -97$nT, $B_y^{\mathrm{DC}} = -44$nT in (a) and $B_x^{\mathrm{DC}} = 14$nT in (b). 
     The half width at half maximum of the resonances in (a) equals $\Gamma/\gamma$, which was fitted for each resonance in (a) and (b) and was found to be within the range 17~($\pm$1)~nT.
     }
    \label{fig:TE_sweep_z}
\end{figure*}

\begin{figure*}
    \centering
    \begin{subfigure}[b]{0.49\linewidth}
        \centering
        \includegraphics[width=\linewidth]{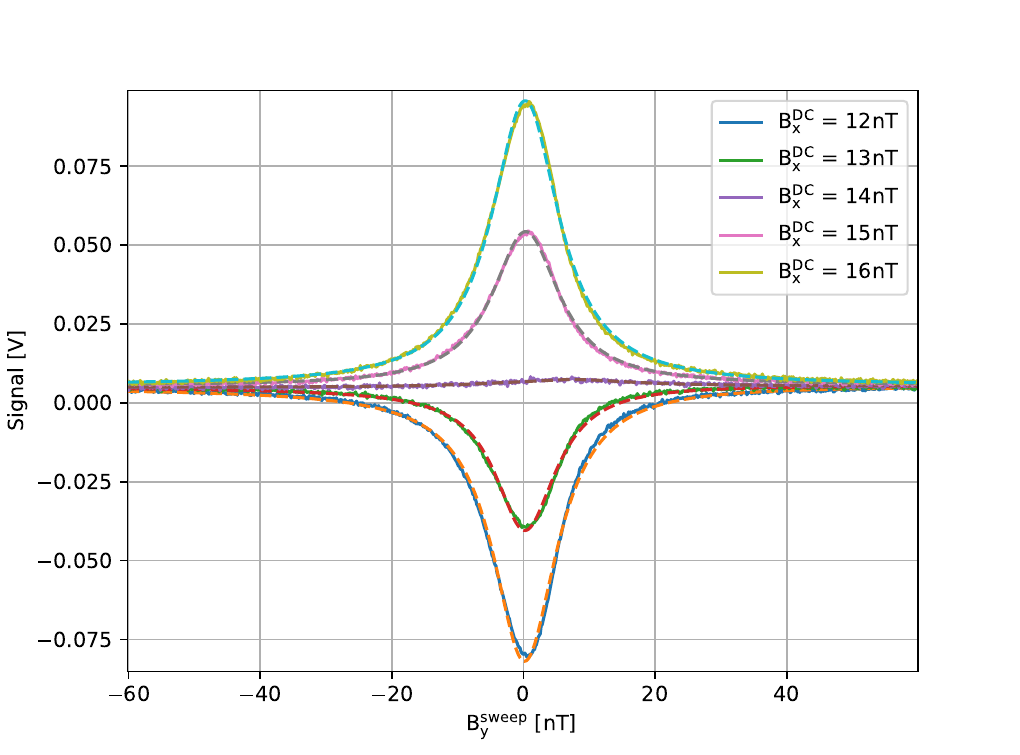}
        \caption{}
        \label{fig:sweep_y_vary_x}
    \end{subfigure}
    \hfill
    \begin{subfigure}[b]{0.49\linewidth}
        \centering
        \includegraphics[width=\linewidth]{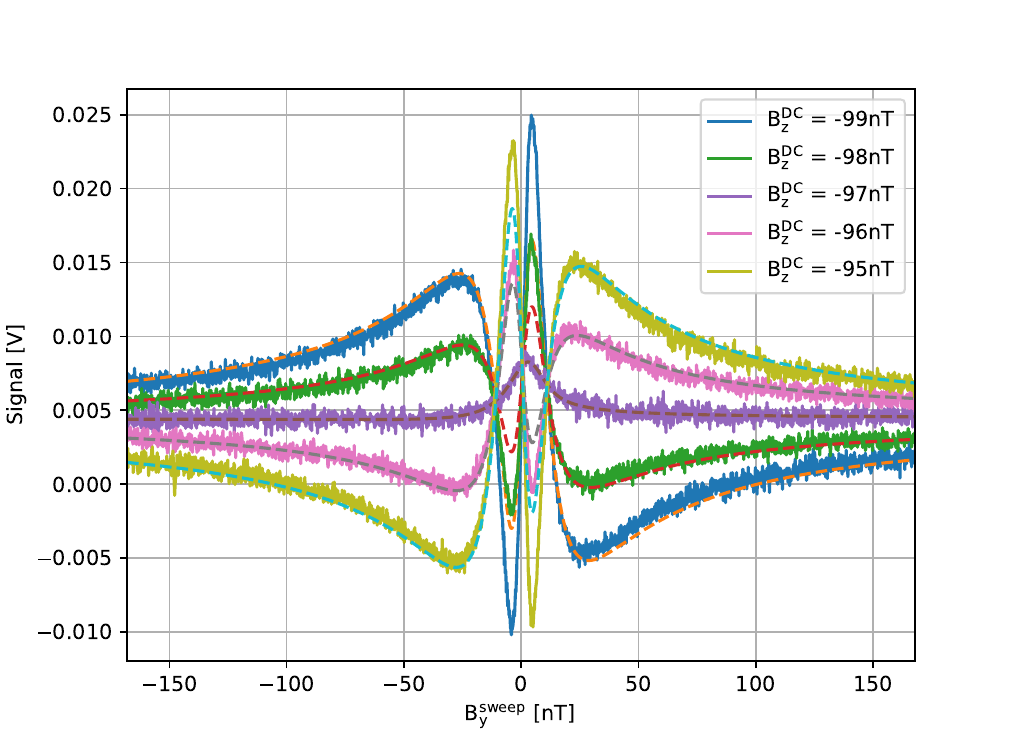}
        \caption{}
        \label{fig:sweep_y_vary_z}
    \end{subfigure}

\begin{subfigure}[b]{0.49\linewidth}
        \centering
        \includegraphics[width=\linewidth]{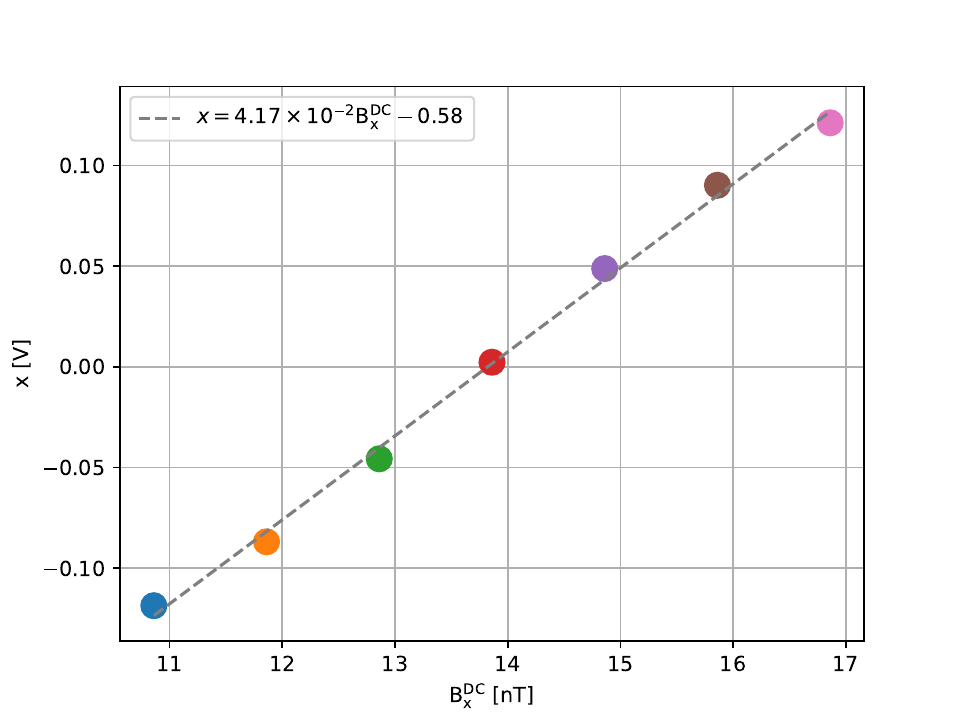}
        \caption{}
        \label{fig:sweep_y_vary_x_fit}
    \end{subfigure}
    \hfill
    \begin{subfigure}[b]{0.49\linewidth}
        \centering
        \includegraphics[width=\linewidth]{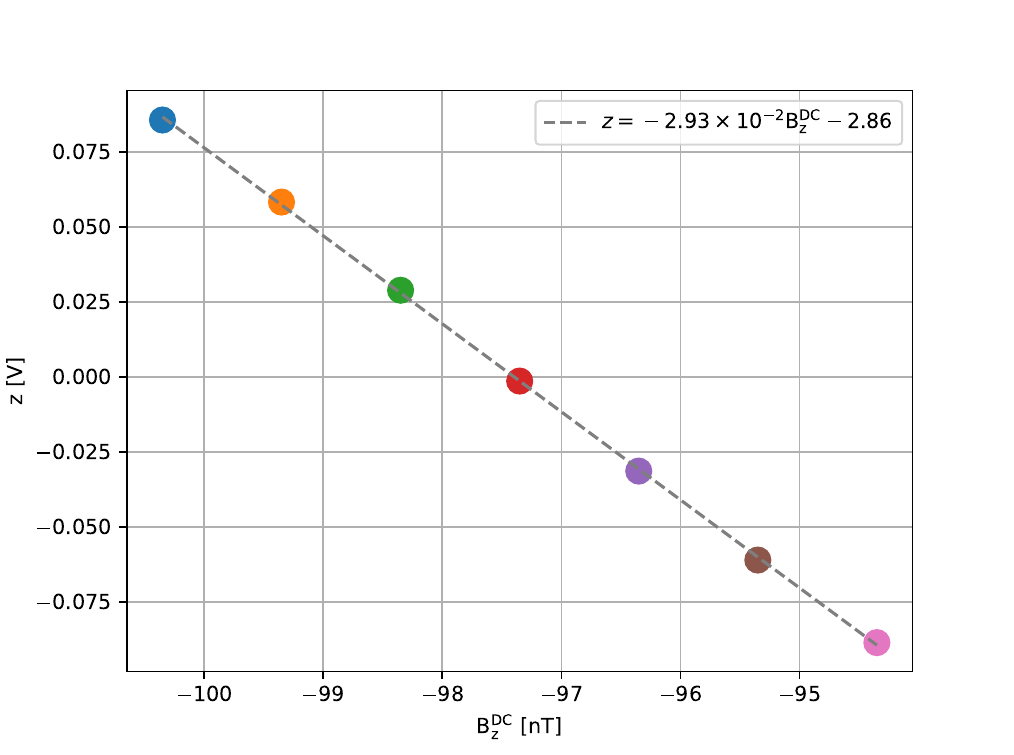}
        \caption{}
        \label{fig:sweep_y_vary_z_fit}
    \end{subfigure}
    
    \caption{Sweeping a magnetic field $B_y^{\mathrm{sweep}}$ applied along the $\ey$ direction near zero-field condition. Solid lines are experimental data, and dashed lines are fits to Eq.~(\ref{eq:S_sweepBy}). (a) and (b) show signals for various settings of an static magnetic field applied along the $\ex$ and $\ez$ directions, respectively. 
    (c) and (d) show the corresponding fit parameters  
    $x\propto \left( B_x^{\mathrm{residual}} + B_x^{\mathrm{DC}} \right)$
    and 
    $z\propto \left( B_z^{\mathrm{residual}} + B_z^{\mathrm{DC}} \right)$. The DC fields are $B_y^{\mathrm{DC}} = -44$nT, $B_z^{\mathrm{DC}} = -97$nT in (a) and $B_x^{\mathrm{DC}} = 14$nT in (b). 
    The parameter $\Gamma/\gamma$ was fitted for each resonance and was found to be within the range
    12 ($\pm$ 1)~nT for the resonances in (a) and 15 ($\pm$1)~nT for the resonances in (b).}   
    \label{fig:TE_sweep_y}
\end{figure*}

We now experimentally demonstrate how to utilize the polarization rotation signal $S$ to null the total static magnetic field $\vec{B}^{\mathrm{residual}}+\vec{B}^{\mathrm{DC}}$. This involves systematically sweeping magnetic fields $\vec{B}^{\mathrm{sweep}}$ applied in the $\ey$ and $\ez$ directions, and then changing the applied static field $\vec{B}^{\mathrm{DC}}$ until a zero signal $S\approx 0$ is obtained. In that case, the total static field should be $\vec{B}^{\mathrm{residual}}+\vec{B}^{\mathrm{DC}} \approx 0$ as seen by Eqs.~(\ref{eq:S_sweepBy}) and (\ref{eq:S_sweepBz}). 
The nulling procedure can be done iteratively, and the results presented here were all recorded in conditions close to zero total static magnetic field.

Figure~\ref{fig:sweep_z_vary_x} presents the measured signals with a linearly sweeping applied magnetic field 
$B^{\mathrm{sweep}}_{z}$ in the $\ez$ direction. 
For these measurements, the static field component $B_{x}^{\mathrm{DC}}$ was varied within the range of 12~nT to 16~nT, while the other components were fixed at $B_{y}^{\mathrm{DC}}=-44$~nT and $B_{z}^{\mathrm{DC}}=-97$~nT. These settings correspond to a total static magnetic field close to zero.
The observed signals looks like Lorentzian lineshapes as predicted by our model, see Eq.~(\ref{eq:S_sweepBz}), when $x \neq 0$ and 
$y = 0$.
Note that for $B_x^{\mathrm{DC}} = 14$~nT, the signal transforms into a flat line as predicted by Eq.~(\ref{eq:S_sweepBz}) when $x = 0$ and $y = 0$. For this setting, the applied static magnetic field precisely cancels the residual magnetic field leading to a net zero static magnetic field $\vec{B}^{\mathrm{residual}}+\vec{B}^{\mathrm{DC}} \approx 0$.

Measurements of the signal while sweeping the applied field $B^{\mathrm{sweep}}_{z}$  in the $\ez$ direction are also obtained when varying the applied static field $B^{\mathrm{DC}}_{y}$ component. Those results are presented in Fig.~\ref{fig:sweep_z_vary_y}. The lineshapes now look dispersive, as predicted by Eq.~(\ref{eq:S_sweepBz}) with $x = 0$ and $y \neq 0$.

The obtained data in Figs.~\ref{fig:sweep_z_vary_x} and \ref{fig:sweep_z_vary_y} are fitted to Eq.~(\ref{eq:S_sweepBz}).  
Figure~\ref{fig:sweep_z_vary_x_fit} shows the parameter 
$x\propto \left( B_x^{\mathrm{residual}} + B_x^{\mathrm{DC}} \right)$ 
obtained by fitting the data in Fig.~\ref{fig:sweep_z_vary_x}. As expected, we see a linear relationship between the applied static field and the total static field. The parameter $y$ (not shown) was also fitted for the data in Fig.~\ref{fig:sweep_z_vary_x} and was found to be much smaller than $x$.
Figure~\ref{fig:sweep_z_vary_y_fit} shows the parameter 
$y\propto \left( B_y^{\mathrm{residual}} + B_y^{\mathrm{DC}} \right)$ 
obtained for the data in Fig.~\ref{fig:sweep_z_vary_y}. Again we see a linear relationship. The parameter $x$ (not shown) was also fitted for the data in Fig.\ref{fig:sweep_z_vary_y} and was found to be much smaller than $y$.

The procedure is now repeated but with instead sweeping a field $B_y^{\mathrm{sweep}}$ applied in the $\ey$  direction. Results are shown in Fig.~\ref{fig:TE_sweep_y}. We see that it is possible to to adjust the $B_{x}^{\mathrm{DC}}$ and $B_{z}^{\mathrm{DC}}$ magnetic field components until the measured signal reaches zero. As mentioned, the nulling procedure can be repeated iteratively, until one obtains a signal very close to zero, for some specific settings of $B_{x}^{\mathrm{DC}}$, $B_{y}^{\mathrm{DC}}$, and $B_{z}^{\mathrm{DC}}$.
The total static field has then been nulled to better than a small fraction of a nT, as seen in Figs.~\ref{fig:sweep_z_vary_x_fit}, \ref{fig:sweep_z_vary_y_fit}, \ref{fig:sweep_y_vary_x_fit} and  \ref{fig:sweep_y_vary_z_fit} which is much smaller than the linewidth of the magnetic resonances, and therefore sufficient for zero-field magnetometry. A larger range where $B^{\mathrm{residual}} + B^{\mathrm{DC}}$ is varied from -40~nT to +40~nT can be seen in Appendix~C in the Supplemental Material where the data is fitted to Eq.~(\ref{eq:Sy_ex}).

\subsection{Zero-field resonances measured by absorption}
\label{sec:exp-absorption}
\begin{figure}
\centering
\includegraphics[width=\linewidth]{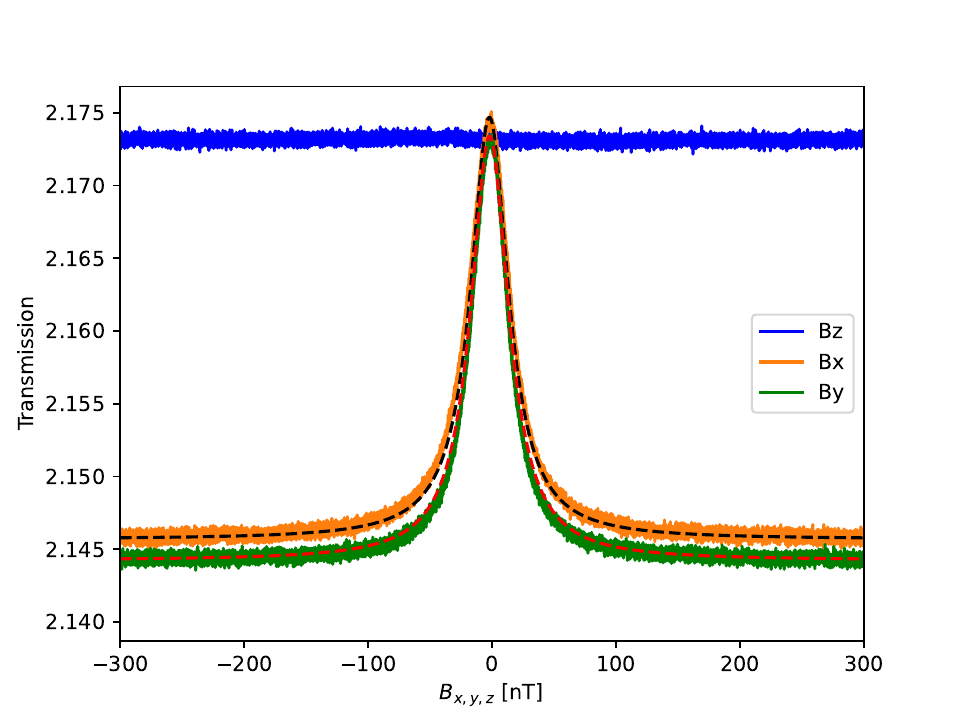}
\caption{When sweeping the magnetic field from positive to negative values for $B_x$, $B_y$, and $B_z$, zero-field resonances can be observed in a direct measurement of the transmitted light intensity. 
The parameter $\Gamma/\gamma$ corresponding to the half width at half maximum of the resonances was fitted and found to be 
18.44~($\pm$ 0.01)~nT and 18.48~($\pm$ 0.01)~nT when sweeping $B_x$ and $B_y$, respectively.
}
\label{fig:Hanle} 
\end{figure}

Having nulled the total  static magnetic field, we now measure the zero-field resonances using absorption, i.e. by measuring the intensity of the transmitted light using a single detector placed immediately after the vapour cell.
If a magnetic field is applied, it affects the quantum state of the caesium atoms as calculated in Eq.~(\ref{eq:zerofieldss}) and will lead to less light being transmitted through the vapour cell. To first order, the atoms are only sensitive to the magnetic field components $B_x$ and $B_y$ which are transverse to the light propagation direction.
Figure \ref{fig:Hanle} shows the sequential sweeping of the three components of the magnetic field using triangular ramps of ±300 (nT).
We see that the signal is constant when sweeping $B_z$, and that the signal has a Lorentzian lineshape when sweeping either $B_x$ or $B_y$, as expected from Eqs.~(\ref{eq:kappa_sweepBx}), (\ref{eq:kappa_sweepBy}) and (\ref{eq:kappa_sweepBz}).
Due to the constant / Lorentzian lineshapes, a direct measurement of the intensity  of the transmitted light   
is not useful for determining the magnetic field components,
as several field values, e.g. $B_x$~\&~$-B_x$, or  $B_y$~\&~$-B_y$ give rise to the same transmitted light intensity, which means that one cannot uniquely infer a field component from the signal.
However, we note that it is possible to measure all components of the magnetic field using modulation techniques, where several auxiliary oscillating magnetic fields are applied \cite{Beato2018,Bertrand2021}.

\begin{figure*}
    \centering
    
    \begin{subfigure}[b]{0.49\linewidth}
        \centering
        \includegraphics[width=\linewidth]{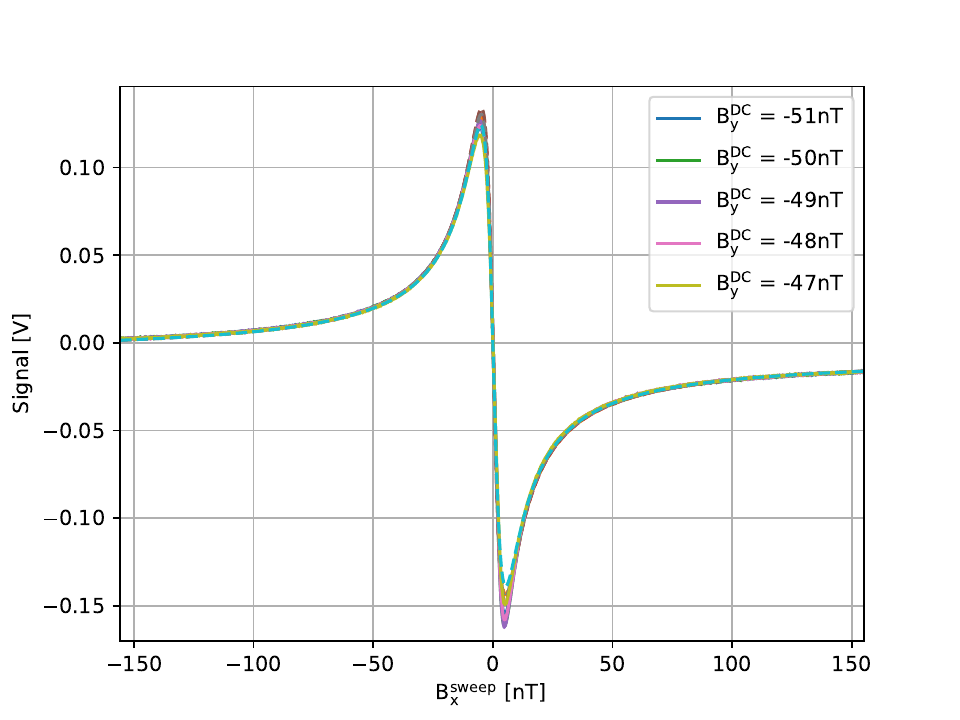}
        \caption{}
        \label{fig:sweep_x_vary_y}
    \end{subfigure}
    \hfill
    \begin{subfigure}[b]{0.49\linewidth}
        \centering
        \includegraphics[width=\linewidth]{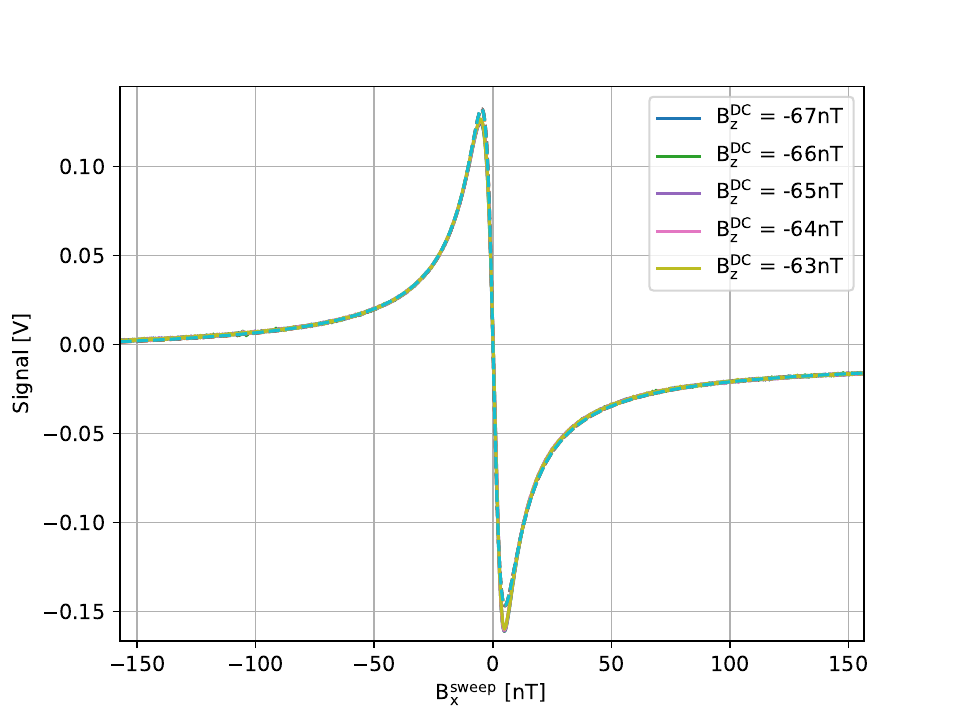}
        \caption{}
        \label{fig:sweep_x_vary_z}
    \end{subfigure}
    
    \caption{Sweeping a magnetic field applied along the $\ex$ direction near zero-field condition. Solid lines are experimental data, and dashed lines are fits to Eq.~(\ref{eq:S_sweepBx}). (a) and (b) show signals for various settings of an magnetic field applied along the $\ey$ and $\ez$ directions, respectively. In (a) $B_z^{\mathrm{DC}} = -65$nT and in (b) $B_y^{\mathrm{DC}} = -49$nT. 
    The parameter $\Gamma/\gamma$ was fitted for each resonance in (a) and (b) and was found to be within the range 11~($\pm$ 1)~nT.
    }
    \label{fig:TE_sweep_x}
\end{figure*}

\subsection{Zero-field magnetometry measured by polarization rotation}

\subsubsection{Magnetic resonance}
We now continue with demonstrating zero-field magnetometry by detecting the change in linear polarization of the transmitted light. As mentioned, this is done using polarimitry and a balanced photodetector.
After having nulled the total  static magnetic field, we  apply a linearly sweeping magnetic field in the $\ex$ direction $B_x^{\mathrm{sweep}}$ with strength ranging from -150~nT to 150~nT and measure the signal for various applied $B^{\mathrm{DC}}_{y}$ and $B^{\mathrm{DC}}_{z}$ static fields. These measurements are presented in Figs.~\ref{fig:sweep_x_vary_y} and \ref{fig:sweep_x_vary_z}. As seen, the signal exhibits a dispersive lineshape and is more or less unaffected by variations in the static magnetic field components as predicted by our theoretical model, see Eq.~(\ref{eq:S_sweepBx}). I.e., in zero-field conditions, the magnetometer is only sensitive to changes in the $\ex$ component of the magnetic field.

\subsubsection{Bandwidth}
We now stop sweeping any magnetic fields.
The bandwidth of the zero-field magnetometer is then assessed by sequentially applying 14 different sinusoidal magnetic fields  with constant amplitude 3.16~$\mathrm{nT_{rms}}$ and varying frequency ranging from 8~Hz to 1818~Hz. Two additional measurements were conducted in which the sinusoidal field was not applied (denoted ``No RF'') and where the balanced photodetector was disconnected from the data-acquisition card (denoted ``Electronic Noise Floor''). These two measurements were performed to assess the noise level of the magnetometer and of the electronic noise of the data acquisition card. For each of the $14+2=16$ measurements, the power spectral density of the signal was calculated and plotted in Fig.~\ref{fig:bandwidth}. Each of the 14 spectra have a sharp peak at the frequency corresponding to the respective oscillation. Those peak heights were fitted as a function of frequency to the response of a first order low-pass filter, i.e.~the function $K/\sqrt{1+\left(f/f_c \right)^2}$.  The low-pass filter response function has been derived for another type of zero-field OPM based on a spin-oriented atomic ensemble \cite{jensen2018magnetocardiography} but is here used phenomenologically to describe the response of our alignment based magnetometer. 
The fit allows us to extract the cut-off frequency 
$f_c = \left( 97 \pm  3 \right)$~Hz
which is the magnetometer bandwidth.
We would expect this number to equal the half width at half maximum of the magnetic resonances in Fig.~\ref{fig:TE_sweep_x} which is 11~nT corresponding to 39~Hz (when multiplying with the gyromagnetic ratio), however, we do experimentally find that the numbers differ by a factor of 2-3.

\subsubsection{Sensitivity}
\begin{figure}
\centering
\includegraphics[width=\linewidth]{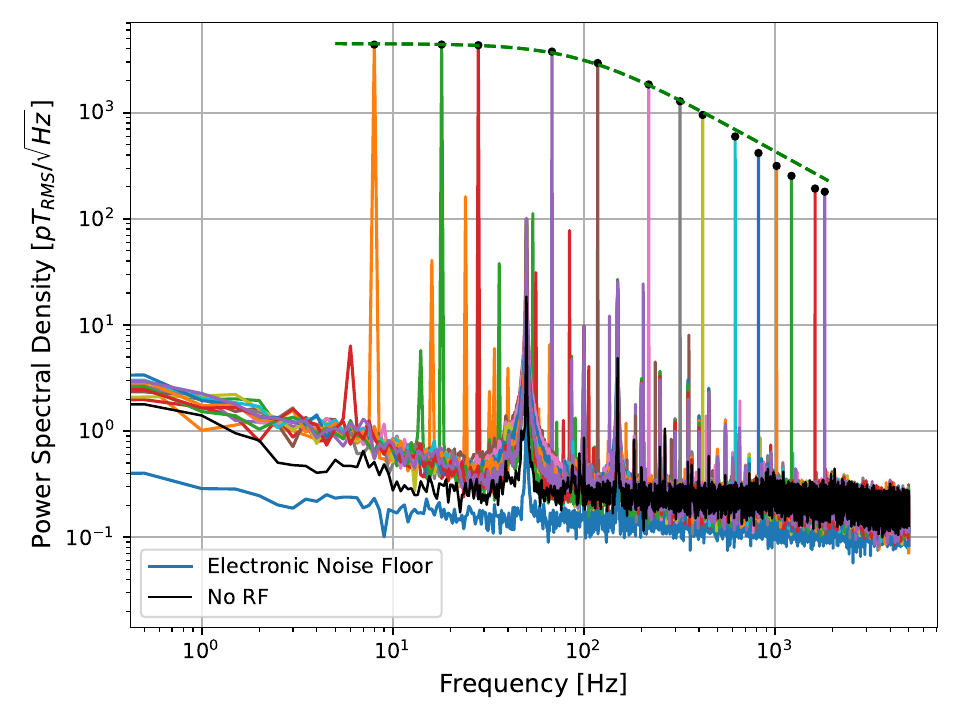}
\caption{Calculated power spectral density (averaged 10 times) using $10 \times 2$~s time traces of the magnetometer signal for 14 different measurements where sinusoidal magnetic fields were applied with various frequencies ranging from 8~Hz to 1818~Hz, and for two additional measurements without a sine wave applied (``No RF''), and with the photodetector disconnected from the data acquisition card (``Electronic Noise Floor'').}
\label{fig:bandwidth}
\end{figure}

The noise floor of the magnetometer obtained with ``No RF'' depends on frequency as seen from Fig.~\ref{fig:bandwidth}. It is $\approx2$~$\mathrm{pT}/\sqrt{\mathrm{Hz}}$ at 1 Hz and then improving to $\approx0.3$~$\mathrm{pT}/\sqrt{\mathrm{Hz}}$ between 10~Hz and 100 Hz.
Our experimentally obtained sensitivity is mainly limited by the electronic noise of our balanced photodetector (data not shown) together with a contribution from the electronic noise of the data acquisition card.

The sensitivity of an optically pumped magnetometer is fundamentally limited by quantum fluctuations in the atomic spin and the light shot noise such that the total quantum noise equals
$\delta \mathrm{B_{quantum}} = \sqrt{\delta \mathrm{B^2_{spin}} + \delta \mathrm{B^2_{shot}}}$.
The spin-projection noise can be written as \cite{ledbetter2007detection}
\begin{equation}
    \delta \mathrm{B_{spin}} = \frac{2 \hbar}{g_F \mu_B \sqrt{nVT_2}},
\end{equation}
where $\mu_B$ is the Bohr magneton, $g_F$ = 1/4 for the $F = 4$ caesium ground state, $n \approx 2.2 \times 10^{16} \mathrm{m^{-3}} $ is the number density of caesium atoms at room temperature (18.5$^{\circ}$C) \cite{rushton2023alignment} and $V = (5 \mathrm{mm})^3$ is the volume of our vapor cell. $T_2$ is the transverse relaxation time and is calculated as
$T_2=1/\Gamma \approx 1/ \left[ 2\pi( 11 \mathrm{nT}\cdot 3.5 \mathrm{Hz/nT}) \right] \approx 4.1$~ms, where we used the $\Gamma/\gamma \approx 11$~nT fitted in   Fig.~\ref{fig:TE_sweep_x} and the gyromagnetic ratio for caesium which can be expressed as $3.5$~Hz/nT.
Ideally, an optically pumped magnetometer has $\delta \mathrm{B_{shot}} = \delta \mathrm{B_{spin}}$ such that 
the total quantum noise 
$\delta \mathrm{B_{quantum}} = \sqrt{2}\delta \mathrm{B_{spin}}$.
Hence the quantum noise limit for our magnetometer is $\delta \mathrm{B_{quantum}} \approx 40 ~ \mathrm{fT/\sqrt{Hz}}$.

\subsubsection{Synthetic magnetocardiography signal}
\begin{figure}
\centering
\includegraphics[width=\linewidth]{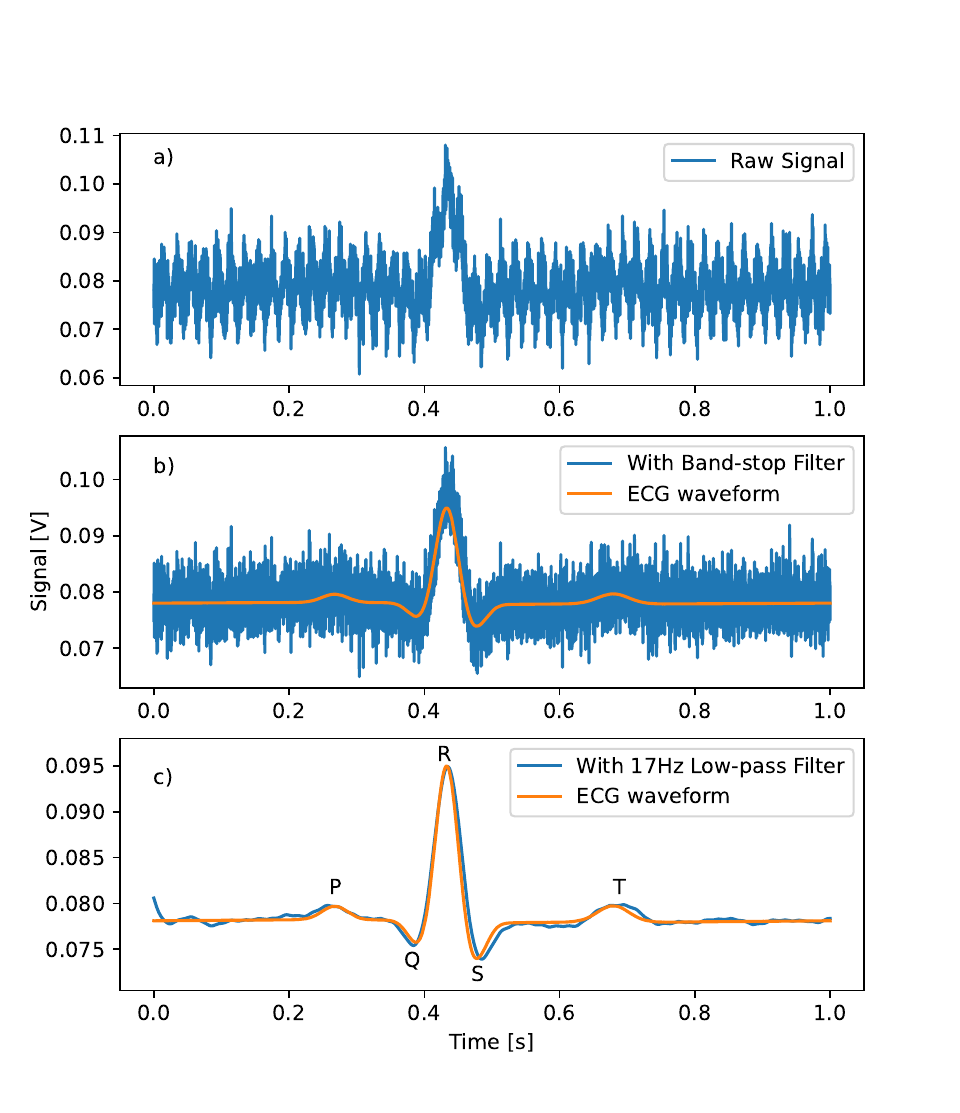}
\caption{Magnetometer time-trace with a synthetic cardiogram waveform applied. (a) Raw data, (b) data processed with a 49-51~Hz bandpass filter, and (c) with additionally a 17~Hz low-pass filter.}
\label{fig:Individual Cardiac Pulse}
\end{figure}

In order to determine if the performance of the magnetometer meets the standards for biomagnetic measurements, particularly magnetocardiography (MCG), a synthetic cardiac signal is applied in the $\ex$ direction with an amplitude of 100~pT peak-to-peak. This amplitude corresponds to the magnetic field of an adult person's heart  \cite{Bison2009apl, Sander2020}
The waveform was generated using the stored  electrocardiogram (ECG) waveform in a function generator (RIGOL DG1032Z) connected to one of the coils.
Figure~\ref{fig:Individual Cardiac Pulse}a shows the raw magnetometer signal, which is not averaged, and was recorded with a 10~kHz sampling rate. The signal is relatively noisy, but simple signal post-processing done by first a 49-51~Hz bandstop filter (to block 50~Hz noise), see Fig.~\ref{fig:Individual Cardiac Pulse}b, and then a 17~Hz low-pass filter (to block high-frequency noise), see Fig.~\ref{fig:Individual Cardiac Pulse}c, significantly improves the signal quality.
We see that the filtered magnetometer signal looks very similar to the  ECG waveform from the function generator. In particular, the real-time measurements captured by the magnetometer reveal the presence of the QRS complex and P and T waves which are features in the cardiogram which are important for medical doctors when assessing the health of a heart.

\section{CONCLUSIONS}

In this study, we investigate the use of spin-aligned atomic ensembles for zero-field magnetometry. A novel method is introduced, which involves assessing the polarization rotation in transmitted light through an atomic vapor. We derive analytical expressions for the generated spin alignment and photodetection signals. By analyzing the obtained polarization rotation signal, we null the magnetic field components through an iterative procedure that adjusts the applied magnetic fields until a zero signal is achieved, indicating the absence of any static magnetic field. 
Achieving a zero-field environment is important for several applications, including nuclear magnetic resonance (NMR) gyroscopes \cite{Walker2016,eklund2021microgyroscope}, and for biomedical measurements, such as magnetoencephalography (MEG) and magnetocardiography (MCG) where non-zero field components leads to artifacts or systematic errors \cite{Borna2022neuro}. 
Our method for nulling the field could potentially be employed in such applications, as demonstrated here by successfully detecting a synthetic cardiac signal. 
As opposed to other methods, e.g.\ using circular polarized light with several modulation fields, our method is modulation-free.
After nulling the field, our magnetometer measures one component of the field using a single laser beam. 
If instead of using a paraffin-coated vapour cell (where all atoms inside the cell are probed), one uses a buffer-gas cell \cite{rushton2023alignment} (where only atoms inside the beam volume are probed),
a vector magnetometer could in principle be implemented using three non-overlapping laser beams. Future work could investigate how to use only one or two laser linearly polarized beams for vector measurements.

We demonstrated a magnetic field sensitivity of $\approx2$~$\mathrm{pT}/\sqrt{\mathrm{Hz}}$ at 1 Hz, which further improved to about 0.3~$\mathrm{pT}/\sqrt{\mathrm{Hz}}$ between 10~Hz and 100~Hz.
The quantum limited sensitivity was calculated to $\delta \mathrm{B_{quantum}} \approx 40 ~ \mathrm{fT/\sqrt{Hz}}$ meaning that further technical improvements are possible. Notably, some commercial optically pumped magnetometers have reached a sensitivity level of around $\sim 20$~fT/$\sqrt{\mathrm{Hz}}$ \cite{Quspin}, however they are operated at an elevated temperature $\sim150^{\circ}$C, as opposed to our magnetometer which was operated at room temperature. To enhance our sensitivity, various strategies can be explored, including the utilization of photodetectors with reduced electronic noise, multiple passes of the laser beam through the vapor cell, employing a larger vapor cell, increasing atom density through vapor cell heating, and using a repump laser beam to transfer atoms from the ``dark'' $F=3$ ground state back to the $F=4$ ground state, which is not probed by the laser light.

\section*{ACKNOWLEDGEMENTS}
This work was supported by 
the QuantERA grant C’MON-QSENS! funded by the Engineering and Physical Sciences Research Council (EPSRC) (Grant No. EP/T027126/1), 
the Novo Nordisk Foundation (Grant No. NNF20OC0064182), 
and the University of Nottingham Hermes Award. Project C’MON-QSENS! is supported by the National Science Centre (2019/32/Z/ST2/00026), Poland under QuantERA, which has received funding from the European Union's Horizon 2020 research and innovation programme under grant agreement no. 731473.

\bibliographystyle{apsrev4-1}

\end{document}


\title{Supplemental Material:\\ Zero field optical magnetometer based on spin-alignment}

\author{A.~Meraki\orcidA{}}
\email{adil.meraki@nottingham.ac.uk}
\affiliation{School of Physics and Astronomy, University of Nottingham, University Park, Nottingham, NG7 2RD, UK}

\author{L.~Elson\orcidB{}}
\affiliation{School of Physics and Astronomy, University of Nottingham, University Park, Nottingham, NG7 2RD, UK}

\author{N.~Ho\orcidE{}}
\affiliation{School of Physics and Astronomy, University of Nottingham, University Park, Nottingham, NG7 2RD, UK}

\author{A.~Akbar\orcidC{}}
\affiliation{School of Physics and Astronomy, University of Nottingham, University Park, Nottingham, NG7 2RD, UK}

\author{M.~Ko\'{z}bia\l\orcidM{}}
\affiliation{Centre for Quantum Optical Technologies, Centre of New Technologies, University of Warsaw, Banacha 2c, 02-097 Warszawa, Poland}

\author{J.~Ko\l{}ody\'{n}ski\orcidJ{}}
\affiliation{Centre for Quantum Optical Technologies, Centre of New Technologies, University of Warsaw, Banacha 2c, 02-097 Warszawa, Poland}

\author{K.~Jensen\orcidD{}} 
\email{kasjensendk@gmail.com}
\affiliation{School of Physics and Astronomy, University of Nottingham, University Park, Nottingham, NG7 2RD, UK}

\maketitle

\appendix

\section{Spin-evolution in matrixform} \label{app:Jxyz}

We can write out the equation of motion for the spin-alignment $\ka=2$ multipoles given by Eq.~(6) in the main text more explicitly as
\begin{equation}
\frac{d}{dt}
\begin{pmatrix} m_{2,-2}\\m_{2,-1}\\m_{2,0}\\m_{2,1}\\m_{2,2}\end{pmatrix}
= 
\left(
-i \omega_x J_x^{(2)}
-i \omega_x J_y^{(2)}
-i \omega_x J_z^{(2)} 
\right)
\begin{pmatrix} m_{2,-2}\\m_{2,-1}\\m_{2,0}\\m_{2,1}\\m_{2,2}\end{pmatrix} 
-\Gamma 
\begin{pmatrix} 
1 & 0 & 0 & 0 & 0 \\
0 & 1 & 0 & 0 & 0 \\
0 & 0 & 1 & 0 & 0 \\
0 & 0 & 0 & 1 & 0 \\
0 & 0 & 0 & 0 & 1 
\end{pmatrix}
\begin{pmatrix} m_{2,-2}\\m_{2,-1}\\m_{2,0}\\m_{2,1}\\m_{2,2}\end{pmatrix}
+ \Gamma
\begin{pmatrix} 0\\0\\m_{2,0}^{\mrm{eq}}\\0\\0\end{pmatrix}.
\end{equation}
The matrices $J_x^{(2)}$, $J_y^{(2)}$, and $J_z^{(2)}$ are the $\ka=2$ representation of the angular momentum operators, which can be written in matrix form as
\begin{equation}
J_x^{(2)} = \hbar
\begin{pmatrix} 
0 & 1 & 0 & 0 & 0 \\
1 & 0 & \sqrt{\frac{3}{2}} & 0 & 0 \\
0 & \sqrt{\frac{3}{2}} & 0 & \sqrt{\frac{3}{2}} & 0 \\
0 & 0 & \sqrt{\frac{3}{2}} & 0 & 1 \\
0 & 0 & 0 & 1 & 0 
\end{pmatrix},
\quad
J_y^{(2)} = \hbar
\begin{pmatrix} 
0 & i & 0 & 0 & 0 \\
-i & 0 & i \sqrt{\frac{3}{2}}  & 0 & 0 \\
0 & -i \sqrt{\frac{3}{2}}  & 0 & i \sqrt{\frac{3}{2}}  & 0 \\
0 & 0 & -i \sqrt{\frac{3}{2}}  & 0 & i \\
0 & 0 & 0 & -i & 0 
\end{pmatrix},
\quad
J_z^{(2)} = \hbar
\begin{pmatrix} 
-2 & 0 & 0 & 0 & 0 \\
0 & -1 & 0 & 0 & 0 \\
0 & 0 & 0 & 0 & 0 \\
0 & 0 & 0 & 1 & 0 \\
0 & 0 & 0 & 0 & 2 
\end{pmatrix}.
\end{equation}

\FloatBarrier

\newpage
\section{Polarization-rotation and light absorption signals in terms of spherical tensor components}

In order to explain the form of Eqs.~(8) and (9) of the main text, 
which describe absorption and polarization-rotation signals, respectively, we identify the symmetry properties that these must fulfil. Symmetry arguments allow us then to determine unambiguously the linear combinations of spherical tensor components, $m_{k q}$ with $k\leq 2$, the two signals must be composed of. As the absorption coefficient $\kappa$ and the polarization-rotation $S$ are linear functions of the density matrix describing the atom \cite{happer1972}, they can be generally written in terms of spherical tensor components as
\begin{align}
    \kappa &= C_0 m_{0,0}+\sum_{q=-1}^1 C_{1q} m_{1q}+\sum_{q=-2}^2 C_{2q} m_{2q}=C_0 m_{0,0}+\mbf C_1^T \mbf{m}_1 + \mbf C_2^T\mbf{m}_2, \label{app:absorption} \\
    S &= D_0 m_{0,0}+\sum_{q=-1}^1 D_{1q} m_{1q}+\sum_{q=-2}^2 D_{2q} m_{2q}=D_0 m_{0,0}+\mbf D_1^T \mbf{m}_1 + \mbf D_2^T\mbf{m}_2. \label{app:rotation}
\end{align}
Here the $C_0$, $C_{1q}$, $C_{2q}$ and the $D_0$, $D_{1q}$, $D_{2q}$ coefficients are complex numbers.
The $C_{1q}$ and $D_{1q}$ coefficients are furthermore organised as row vectors ($\mbf C_1^T$ and $\mbf D_1^T$)
in 3-dimensional space, and the $C_{2q}$ and $D_{2q}$ coefficients are organised as row vectors ($\mbf C_2^T$ and $\mbf D_2^T$) in 5-dimensional space.

Now, rotating the atomic state around the light propagation axis $x$ by $\pi$ must be equivalent to rotating the light polarization by $-\pi$ around $x$. Moreover, since the light is linearly polarised, the latter rotation does not affect the system, so both the absorption and polarization-rotation signal should not be affected by either of the operations. These, acting on $\mbf m_1$ and $\mbf m_2$ respectively, have a matrix form
\begin{align}
\mbf R_{x}^{(1)}=\exp [-i \pi J_x^{(1)} /\hbar]&=\begin{pmatrix} 
0 & 0 & -1 \\
0 & -1 & 0 \\
-1 & 0 & 0
\end{pmatrix}, \\
\mbf R_{x}^{(2)}=\exp [-i \pi J_x^{(2)} /\hbar]&=\begin{pmatrix} 
0 & 0 & 0 & 0 & 1 \\
0 & 0 & 0 & 1 & 0 \\
0 & 0 & 1 & 0 & 0 \\
0 & 1 & 0 & 0 & 0 \\
1 & 0 & 0 & 0 & 0 
\end{pmatrix},
\end{align}
while the $m_{0,0}$ component remains constant. 
The matrices $J_x^{(1)}$ and $J_x^{(2)}$ are the $\ka=1$  and $\ka=2$ representations of the $x$-component of the angular momentum operator, respectively.
Hence, for the signals \eqref{app:absorption} and \eqref{app:rotation} to be invariant we must impose
\begin{align}
    &\mbf C_1^T \mbf{m}_1 + \mbf C_2^T\mbf{m}_2=\mbf C_1^T \mbf R_{x}^{(1)} \mbf{m}_1 + \mbf C_2^T \mbf R_{x}^{(2)} \mbf{m}_2, \\
    &\mbf D_1^T \mbf{m}_1 + \mbf D_2^T\mbf{m}_2=\mbf D_1^T \mbf R_{x}^{(1)} \mbf{m}_1 + \mbf D_2^T \mbf R_{x}^{(2)} \mbf{m}_2
\end{align}
to hold for any $\mbf m_1$, $\mbf m_2$. This means, that $\mbf C_1$ and $\mbf D_1$ must be eigenvectors of $(\mbf R_{x}^{(2)})^T$, while $\mbf C_2$ and $\mbf D_2$ must be eigenvectors of $(\mbf R_{x}^{(1)})^T$, all corresponding to the eigenvalue 1, i.e.
\begin{align}
\mbf C_1, \mbf D_1 \in &  \,\mrm{span}\left\{\begin{pmatrix} 
1 \\
0 \\
-1 
\end{pmatrix}\right\}, \label{app:span1} \\
\mbf C_2, \mbf D_2 \in &  \,\mrm{span}\left\{\begin{pmatrix} 
1 \\
0 \\
0 \\
0 \\
1 
\end{pmatrix},
\begin{pmatrix} 
0 \\
1 \\
0 \\
1 \\
0 
\end{pmatrix},
\begin{pmatrix} 
0 \\
0 \\
1 \\
0 \\
0 
\end{pmatrix}\right\}. \label{app:span2}
\end{align}

Furthermore, the light absorption $\kappa$, for a given direction of light polarization should not depend on the direction of light propagation at all. Therefore, as the rotation of the atomic state around the light polarization direction by a given angle $\varphi$ is given by operators
\begin{align}
\mbf R_{\varphi}^{(1)}=\exp [-i \varphi J_z^{(1)} /\hbar ]&=\begin{pmatrix} 
e^{i\varphi} & 0 & 0\\
0 & 1 & 0 \\
0 & 0 & e^{-i\varphi}
\end{pmatrix} \\
\mbf R_{\varphi}^{(2)}=\exp [-i \varphi J_z^{(2)} /\hbar]&=\begin{pmatrix} 
e^{2i\varphi} & 0 & 0 & 0 & 0 \\
0 & e^{i\varphi} & 0 & 0 & 0 \\
0 & 0 & 1 & 0 & 0 \\
0 & 0 & 0 & e^{-i\varphi} & 0 \\
0 & 0 & 0 & 0 & e^{-2i\varphi} 
\end{pmatrix},
\end{align}
we must analogously impose that $\mbf C_1$ and $\mbf C_2$ must also be eigenvectors of $(\mbf R_{\varphi}^{(1)})^T$ and $(\mbf R_{\varphi}^{(2)})^T$, respectively, corresponding to the eigenvalue 1. Hence
\begin{align}
\mbf C_1\in &  \,\mrm{span}\left\{\begin{pmatrix} 
0 \\
1 \\
0 
\end{pmatrix}\right\}, \\
\mbf C_2\in & \,\mrm{span}\left\{
\begin{pmatrix} 
0 \\
0 \\
1 \\
0 \\
0 
\end{pmatrix}\right\}.
\end{align}
The $m_{0,0}$ component in this case also remains unchanged. By combining the above expressions with \eqref{app:span1} and \eqref{app:span2} we arrive to a conclusion, that $\mbf C_1=\begin{pmatrix}
    0 & 0 & 0
\end{pmatrix}^T$, while $\mbf C_2 \propto \begin{pmatrix}
    0 & 0 & 1 & 0 & 0
\end{pmatrix}^T$. Since there are also no constraints on $C_0$, Eq.~\eqref{app:absorption} takes form of Eq.~(8).

Considering now the polarization-rotation signal $S$, we expect that if the light propagates in the opposite direction, here $-x$, the polarization-rotation has the same magnitude, but the opposite sign, so that, e.g., a clockwise light polarization-rotation around $x$ becomes anti-clockwise. On the other hand, reverting the direction of light propagation is equivalent to rotation of the atomic system around $z$ axis by $\pi$, whose matrix forms read
\begin{align}
\mbf R_{z}^{(1)}=\exp [-i \pi J_z^{(1)} /\hbar]&=\begin{pmatrix} 
-1 & 0 & 0\\
0 & 1 & 0 \\
0 & 0 & -1
\end{pmatrix}, \\
\mbf R_{z}^{(2)}=\exp [-i \pi J_z^{(2)} /\hbar]&=\begin{pmatrix} 
1 & 0 & 0 & 0 & 0 \\
0 & -1 & 0 & 0 & 0 \\
0 & 0 & 1 & 0 & 0 \\
0 & 0 & 0 & -1 & 0 \\
0 & 0 & 0 & 0 & 1 
\end{pmatrix}.
\end{align}
Hence, following the same arguments as before, we must impose that the polarization rotation signal \eqref{app:rotation} fulfils
\begin{equation}
    D_0 m_{0,0}+\mbf D_1^T \mbf{m}_1 + \mbf D_2^T\mbf{m}_2=-(D_0 m_{0,0}+\mbf D_1^T \mbf R_{z}^{(1)} \mbf{m}_1 + \mbf D_2^T \mbf R_{z}^{(2)} \mbf{m}_2)
\end{equation}
for any $m_{0,0}$, $\mbf m_1$ and $\mbf m_2$. As a result, this time we obtain $D_0=0$ and
\begin{align}
\mbf D_1\in & \,\mrm{span}\left\{\begin{pmatrix} 
1 \\
0 \\
0 
\end{pmatrix},\begin{pmatrix} 
0 \\
0 \\
1 
\end{pmatrix}\right\}, \\
\mbf D_2\in & \,\mrm{span}\left\{
\begin{pmatrix} 
0 \\
1 \\
0 \\
0 \\
0 
\end{pmatrix},\begin{pmatrix} 
0 \\
0 \\
0 \\
1 \\
0 
\end{pmatrix}\right\},
\end{align}
which are eigenspaces of $(\mbf R_{z}^{(1)})^T$ and $(\mbf R_{z}^{(2)})^T$ respectively, now corresponding to the eigenvalue~$-1$. Again, by considering also \eqref{app:span1} and \eqref{app:span2}, we arrive to the conclusion that $\mbf D_1 \propto \begin{pmatrix}
    -1 & 0 & 1
\end{pmatrix}^T$ and $\mbf D_2 \propto \begin{pmatrix}
    0 & 1 & 0 & 1 & 0
\end{pmatrix}^T$. The rotation signal reads
\begin{equation}
    S=D_{1,1}(m_{1,1}-m_{1,-1}) + D_{2,1}(m_{2,1}+m_{2,-1})], \label{app:rotation_signal}
\end{equation}
where $D_{2,1}$ needs to be purely imaginary to ensure that the signal is real. Moreover, since only the $m_{2q}$ components can be made non zero when pumping the ensemble with linearly polarised light \cite{happer1972}, one may in our case set $m_{1q}=0$ for all $q$ above, in order to arrive at Eq.~(9) of the main text.

As a side note, let us also observe that since $m_{1,1}=\Tr[T^{1\dag}_1\rho]$, $m_{1,-1}=\Tr[T^{1\dag}_{-1}\rho]=\Tr[-T^{1}_1\rho]$, and also $T^1_1\propto F_+$, the part of the polarization-rotation signal \eqref{app:rotation_signal} for $k=1$, i.e.~the multipoles defining the orientation vector, generally reads
\begin{equation}
    D_{1,1}(m_{1,1}-m_{1,-1})=D_{1,1}\Tr[(T^{1\dag}_1+T^{1}_{1})\rho]\propto\Tr[(F_+ +F_-)\rho]=2\langle F_x \rangle.
\end{equation}
This shows that for orientation-based magnetometers, the polarization rotation signal is proportional to the projection of the angular momentum $\langle F_x \rangle$ along the light propagation direction, which is a well-known result for orientation-based magnetometers.

\newpage

\section{Sweeping the magnetic field over a larger range}

\begin{figure}[H]
    \centering

    \begin{subfigure}[b]{0.49\linewidth}
        \centering
        \includegraphics[width=\linewidth]{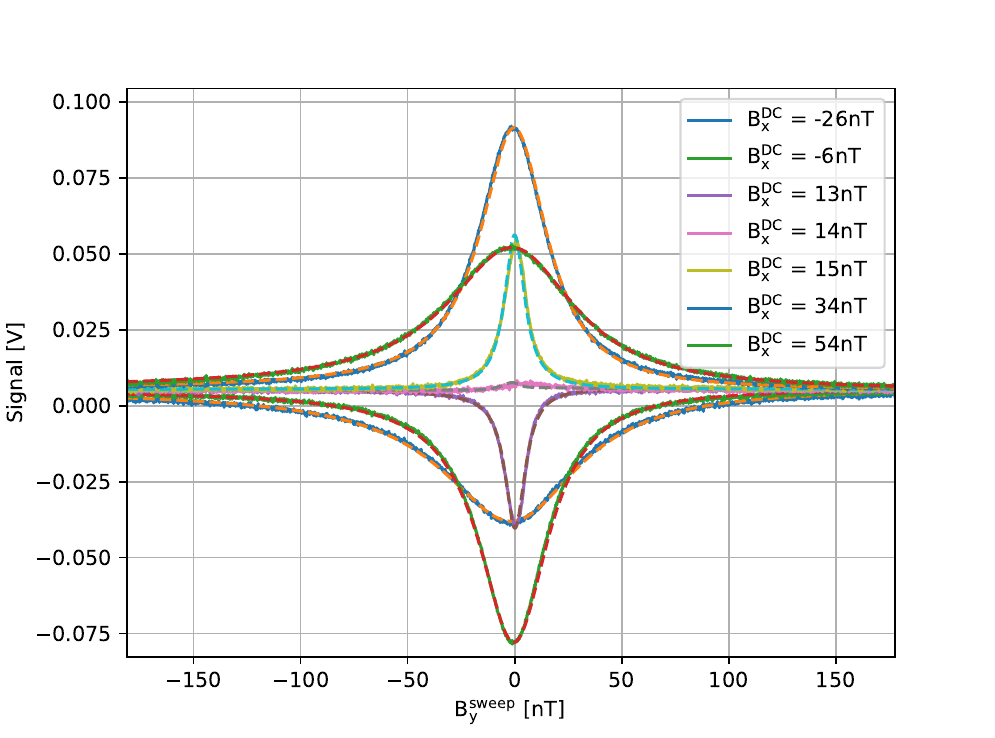}
        \caption{}
        \label{fig:sweep_y_vary_x_full}
    \end{subfigure}
    \hfill
    \begin{subfigure}[b]{0.49\linewidth}
        \centering
        \includegraphics[width=\linewidth]{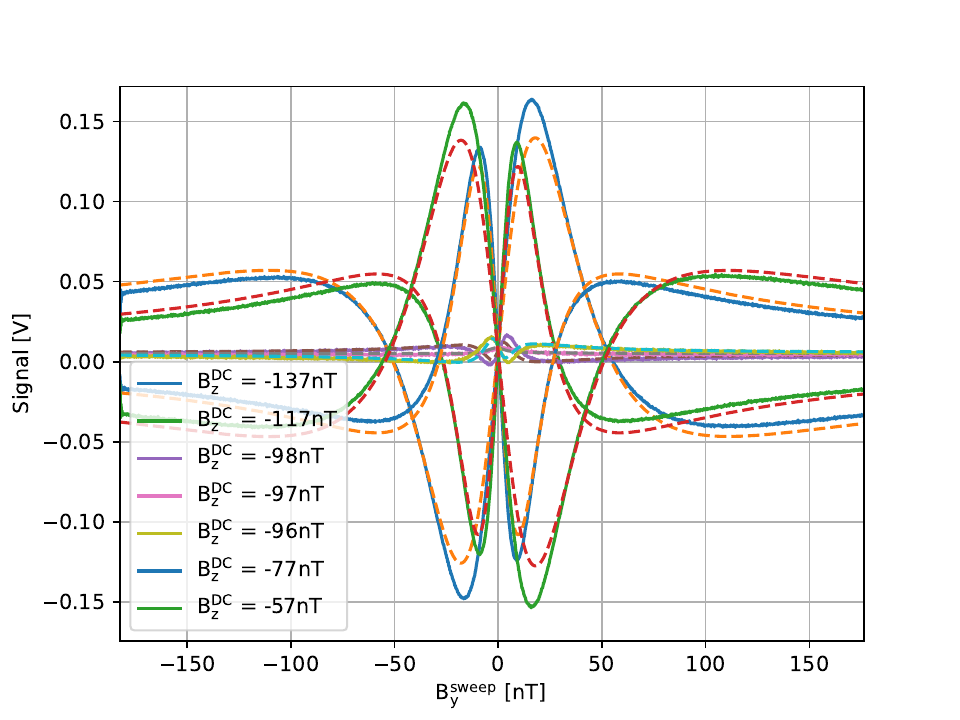}
        \caption{}
        \label{fig:sweep_y_vary_z_full}
    \end{subfigure}
    
    \begin{subfigure}[b]{0.49\linewidth}
        \centering
        \includegraphics[width=\linewidth]{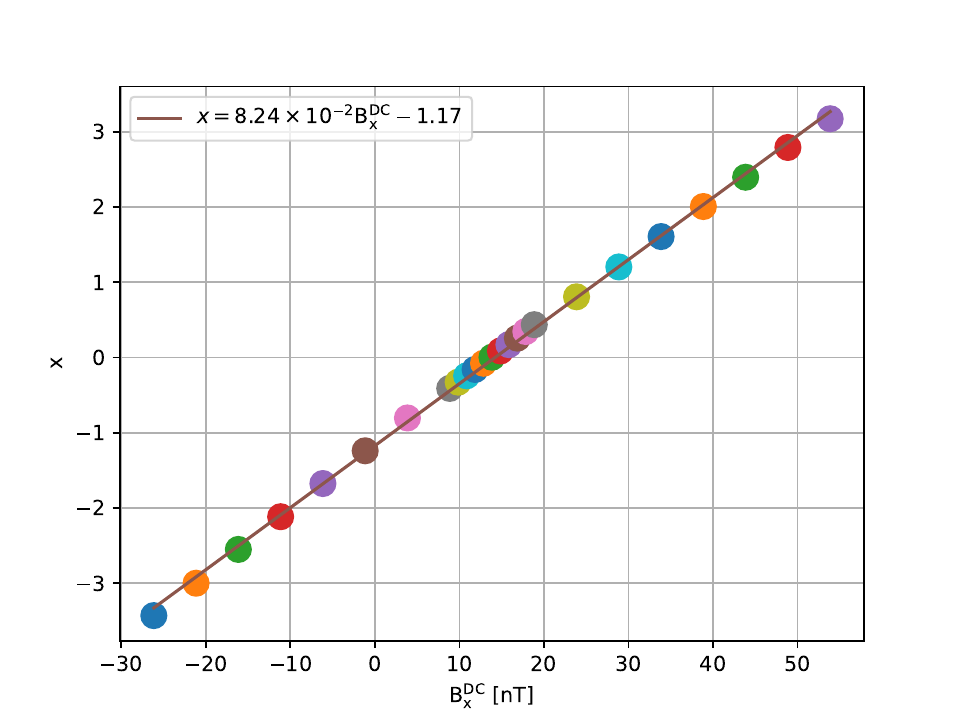}
        \caption{}
        \label{fig:sweep_y_vary_x_BR}
    \end{subfigure}
    \hfill
    \begin{subfigure}[b]{0.49\linewidth}
        \centering
        \includegraphics[width=\linewidth]{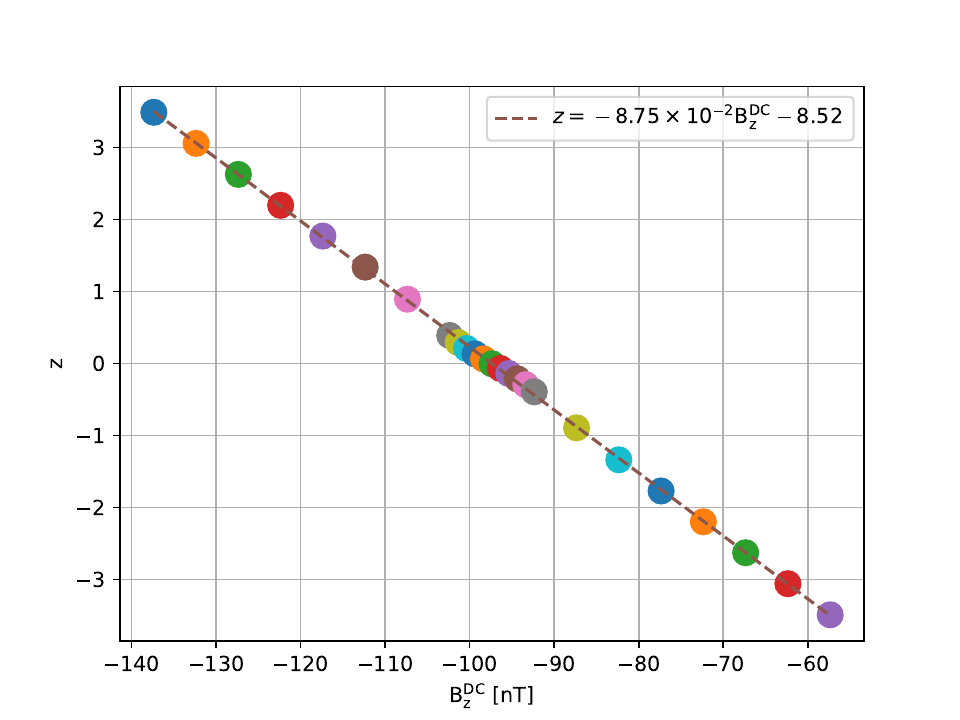}
        \caption{}
        \label{fig:sweep_y_vary_z_BR}
    \end{subfigure}
    
    \caption{Sweeping a magnetic field applied along the $\ey$-direction near zero-field condition. Solid lines are experimental data, and dashed lines are fits to Eq.~(11). (a) and (b) show signals for various settings of an magnetic field applied along the $\ex$ and $\ez$ directions, respectively. 
    (c) and (d) show the corresponding fit parameters  $x\propto B_x$ and $z\propto B_z$ which are proportional to different components of the net magnetic field. The DC fields are $B_y^{\mathrm{DC}} = -44$nT, $B_z^{\mathrm{DC}} = -97$nT in (a) and $B_x^{\mathrm{DC}} = 14$nT in (b). 
    The parameter $\Gamma/\gamma$ was a global fit parameter for all the resonances in (a) and was found to be 11.66$(\pm 0.03)$~nT. Similarly, we found  $\Gamma/\gamma=10.55(\pm 0.03)$~nT for the resonances in (b).    
    }
    \label{fig:full_sweep_y}
\end{figure}

\begin{figure}[H]
    \centering

    \begin{subfigure}[b]{0.49\linewidth}
        \centering
        \includegraphics[width=\linewidth]{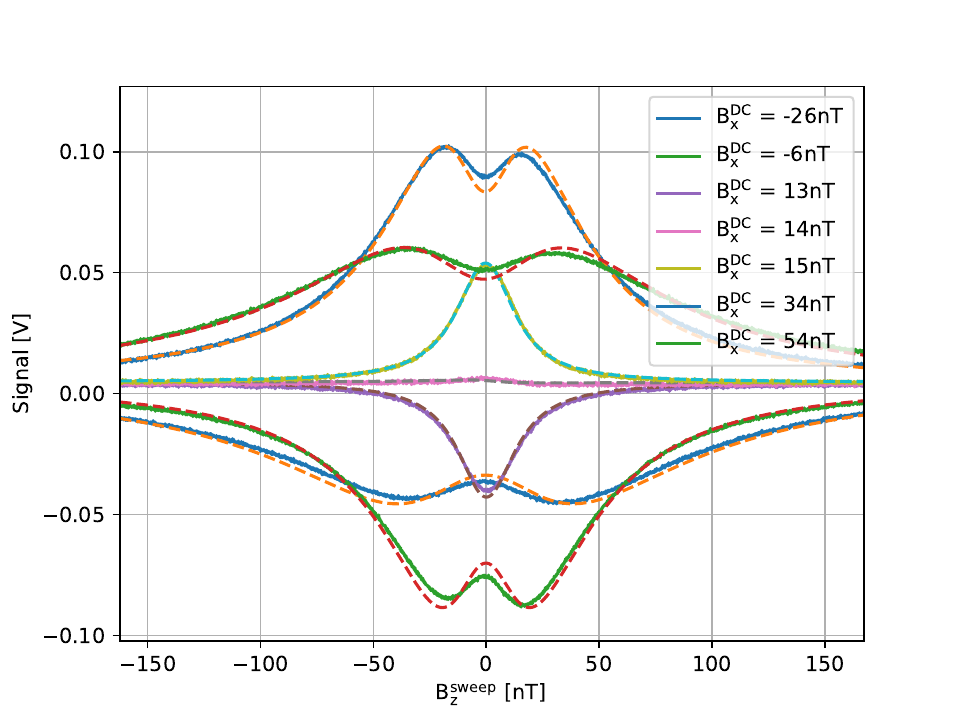}
        \caption{}
        \label{fig:sweep_z_vary_x_full}
    \end{subfigure}
    \hfill
    \begin{subfigure}[b]{0.49\linewidth}
        \centering
        \includegraphics[width=\linewidth]{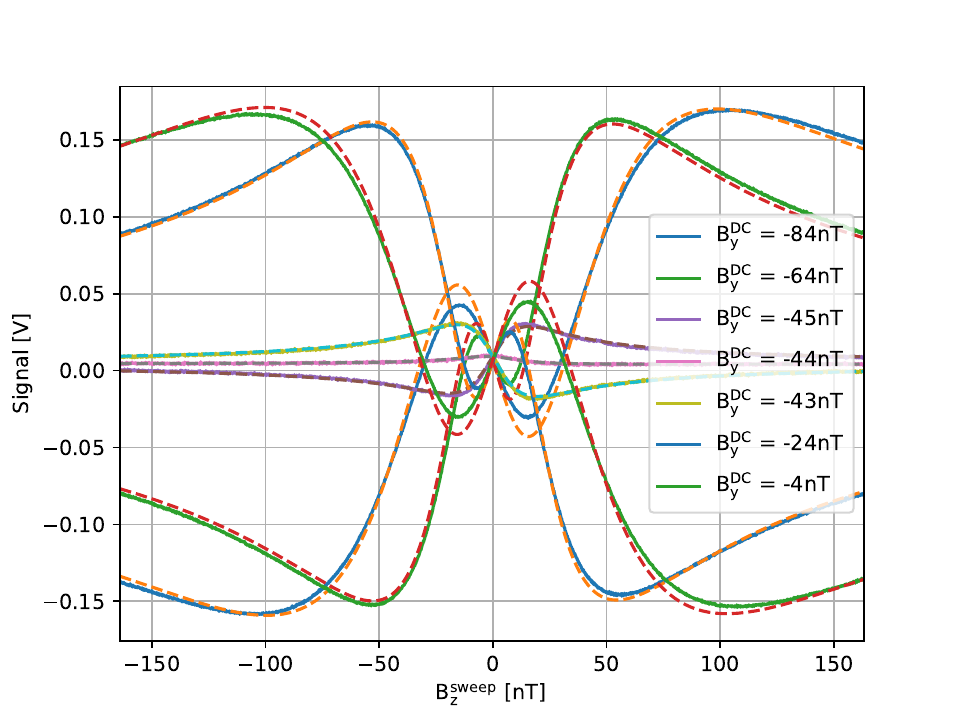}
        \caption{}
        \label{fig:sweep_z_vary_y_full}
    \end{subfigure}
    
    \begin{subfigure}[b]{0.49\linewidth}
        \centering
        \includegraphics[width=\linewidth]{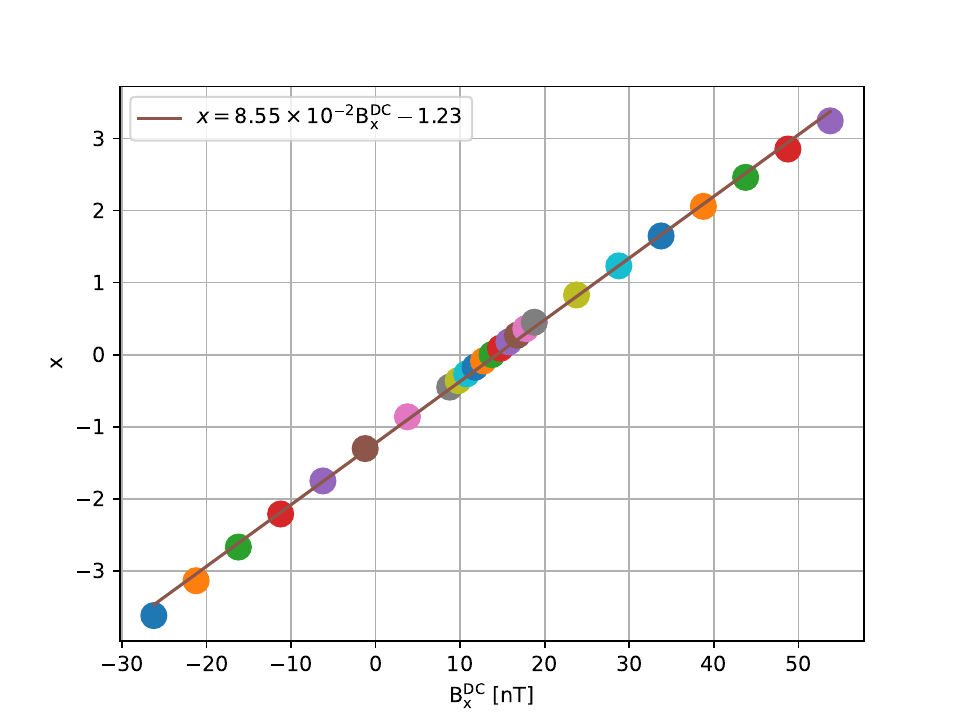}
        \caption{}
        \label{fig:sweep_z_vary_x_BR}
    \end{subfigure}
    \hfill
    \begin{subfigure}[b]{0.49\linewidth}
        \centering
        \includegraphics[width=\linewidth]{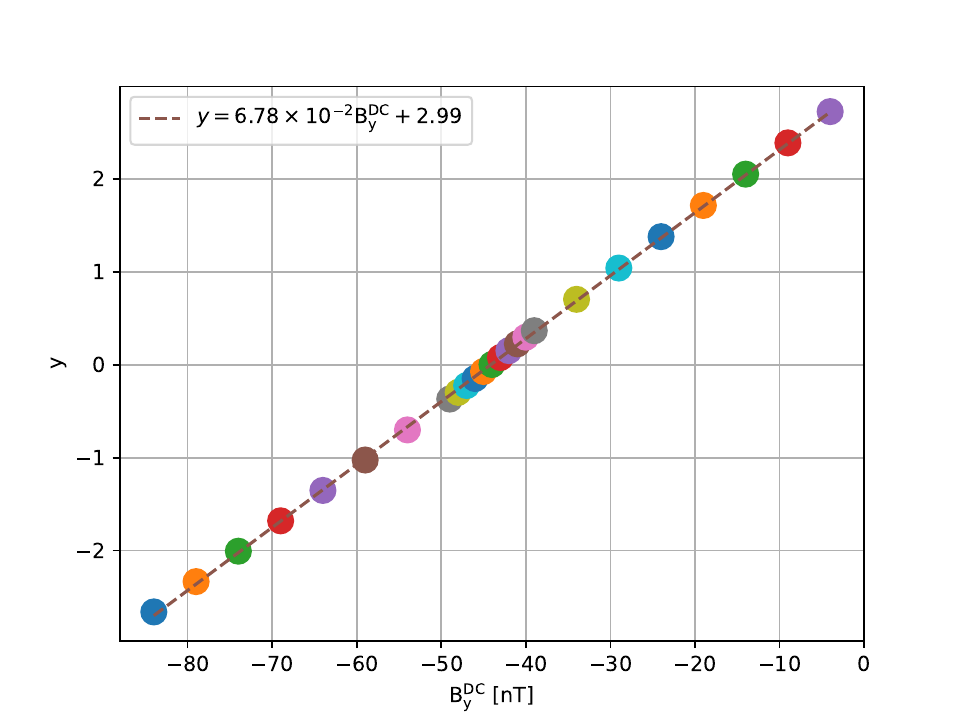}
        \caption{}
        \label{fig:sweep_z_vary_y_BR}
    \end{subfigure}
    
    \caption{Sweeping a magnetic field applied along the $\ez$-direction near zero-field condition. Solid lines are experimental data, and dashed lines are fits to Eq.~(11). (a) and (b) show signals for various settings of an magnetic field applied along the $\ex$ and $\ey$ directions, respectively. 
    (c) and (d) show the corresponding fit parameters  $x\propto B_x$ and $y\propto B_y$ which are proportional to different components of the net magnetic field. The DC fields are  $B_z^{\mathrm{DC}} = -97$nT, $B_y^{\mathrm{DC}} = -44$nT in (a) and $B_x^{\mathrm{DC}} = 14$nT in (b). 
    The parameter $\Gamma/\gamma$ was a global fit parameter for all the resonances in (a) and was found to be 15.88$(\pm 0.03)$~nT. Similarly, we found  $\Gamma/\gamma=17.50 (\pm 0.03)$~nT for the resonances in (b).}
    \label{fig:full_sweep_z}
\end{figure}

Alongside looking at how the polarization rotation signal changes with small changes of the overall field in the magnetic shield we also wanted to study a range that is larger than the linewidth of the magnetometer. The field was varied such that the overall field in each direction was -40~nT to +40~nT. The results were then fitted to Eq.~(11) in the main text. The fitting procedure here is slightly different as the fit parameters for the linewidth and offset are shared between all 25 data sets. Such that when we are sweeping the $\ey$ field only x and z are fitted individually for each data set. Similarly, when sweeping the $\ez$ field only the x and y parameters are not shared between the data sets. Figure~\ref{fig:full_sweep_y} shows the results for sweeping the $\ey$ field. It can be seen that when fitting to the full equation a linear relationship is still seen. Figure~\ref{fig:full_sweep_z} shows a similar relationship for when the $\ez$ field is swept. Hence the theory is not limited to the residual field being very close to zero. 

\bibliographystyle{apsrev4-1}